\documentclass[]{aa}  

\usepackage{graphicx}
\usepackage{txfonts}
\usepackage{soul}
\usepackage{listings}
\usepackage{courier}
\usepackage{cancel}
\usepackage{subcaption}
\usepackage{tikz}

\newcommand{\Gnaught}{G_0}

\usepackage{siunitx}
\DeclareSIUnit{\mSun}{\ensuremath{M_\odot}}
\DeclareSIUnit{\yr}{yr}
\DeclareSIUnit{\erg}{erg}
\DeclareSIUnit{\Gnaught}{\ensuremath{G_0}}
\DeclareSIUnit{\AU}{AU}
\DeclareSIUnit{\parsec}{pc}
\DeclareSIUnit{\Kelvin}{K}

\usepackage[version=4]{mhchem}

\newcommand{\mSun}{M_\odot}

\newcommand{\GravC}{\mathcal{G}}
\newcommand{\sound}{\mathrm{s}}

\newcommand{\mStar}{M_*}

\newcommand{\alphaturb}{\alpha_\mathrm{SS}}
\newcommand{\wind}{\mathrm{DW}}
\newcommand{\alphaDW}{\alpha_\wind}

\newcommand{\gas}{\mathrm{g}}
\newcommand{\disc}{\mathrm{disc}}

\newcommand{\betaT}{\beta_\mathrm{T}}

\newcommand{\betaf}{{\beta_\mathrm{f}}}
\newcommand{\visc}{\mathrm{visc}}
\newcommand{\Kepl}{\mathrm{K}}

\newcommand{\extphot}{\mathrm{evap}}
\newcommand{\St}{\mathrm{St}}
\newcommand{\Sc}{\mathrm{Sc}}
\newcommand{\solid}{\mathrm{s}}
\newcommand{\Rc}{{R_\mathrm{c}}}
\newcommand{\trunc}{\mathrm{t}}
\newcommand{\dust}{\mathrm{d}}
\newcommand{\tot}{\mathrm{tot}}

\newcommand{\psiDW}{\psi_\mathrm{DW}}

\newcommand{\deplete}{\mathrm{dep}}
\newcommand{\tIRex}{t_\mathrm{IR.ex}}
\newcommand{\tMacc}{t_\mathrm{acc.}}
\newcommand{\tMg}{t_{\deplete,\mathrm{gas}}}
\newcommand{\tMd}{t_{\deplete,\mathrm{dust}}}

\usepackage{natbib,twoopt}
\usepackage[breaklinks=true]{hyperref} 
\bibpunct{(}{)}{;}{a}{}{,}             
\makeatletter
  \newcommandtwoopt{\citeads}[3][][]{\href{http://ui.adsabs.harvard.edu/abs/#3}%
    {\def\hyper@linkstart##1##2{}%
     \let\hyper@linkend\@empty\citet[#1][#2]{#3}}}
  \newcommandtwoopt{\citepads}[3][][]{\href{http://ui.adsabs.harvard.edu/abs/#3}%
    {\def\hyper@linkstart##1##2{}%
     \let\hyper@linkend\@empty\citep[#1][#2]{#3}}}
  \newcommandtwoopt{\citeyearads}[3][][]%
    {\href{http://ui.adsabs.harvard.edu/abs/#3}
    {\def\hyper@linkstart##1##2{}%
     \let\hyper@linkend\@empty\citeyear[#1][#2]{#3}}}
  \newcommandtwoopt{\citealtads}[3][][]%
    {\href{http://ui.adsabs.harvard.edu/abs/#3}
    {\def\hyper@linkstart##1##2{}%
     \let\hyper@linkend\@empty\citealt[#1][#2]{#3}}}
\makeatother

\begin{document}

    \title{Timescales diagnostics for saving viscous and MHD-driven dusty discs from external photoevaporation}

    \titlerunning{Viscous and MHD-driven dusty discs with external FUV}

    \subtitle{}

    \author{G. Pichierri
            \inst{1}
            \and
           G. Rosotti\inst{1}
           \and 
           R. Anania\inst{1,2}
           \and
           G. Lodato\inst{1}
          }

    \institute{Dipartimento di Fisica, Università degli Studi di Milano, Via Celoria 16, I-20133 Milano, Italy;\\
            \email{gabriele.pichierri@unimi.it}
            \and
            School of Physics, Trinity College Dublin, the University of Dublin, College Green, Dublin 2, Ireland
            \\
            }

    \date{...; }
 
    \abstract
    {
    The evolution of protoplanetary discs is a function of their internal processes and of the environment in which these discs find themselves. It is unclear if angular momentum is mainly removed viscously or by magnetic winds, or by a combination of the two. While external photoevaporation is expected to severely influence disc evolution and eventually dispersal, there are observational limitations towards highly irradiated discs. The interplay between these ingredients and their combined effects on the gas and dust distributions within the disc are poorly understood.
    }
    {
    We investigate, for the first time, the evolution of both the gaseous and solid components of viscous, MHD-wind or hybrid discs, in combination with external FUV-driven mass loss. We test which combinations of parameters may protect discs from the external irradiation, allowing the solid component to live long enough to allow planet formation to succeed.
    }
    {
    We run a suite of 1D simulations of smooth discs with varying initial sizes, different levels of viscous and MHD-wind stresses modelled via an $\alpha$ parametrisation, and strengths of the external FUV environment. We then track disc properties such as their radii, various lifetime diagnostics, and the amount of dust removed by the photoevaporative wind, as a function of the underlying parameters.
    }
    {
    We find that the biggest role in determining the fate of discs is played by a combination of a disc's ability to spread radially outwards and the strength of FUV-driven erosion. While MHD wind-driven discs experience less FUV erosion due to the lack of spread, they do not live for longer compared to viscously evolving discs, especially at low-to-moderate FUV fluxes, while higher fluxes ($\gtrsim 100\,G_0$) yield disc lifetimes that are rather insensitive to the disc's angular momentum transport mechanism. Specifically, for the solid component, the biggest role is played by a combination of inward drift and removal by FUV winds. This points to the importance of other physical ingredients, such as disc substructures, even in highly-irradiated disc regions, in order to retain solids.
    }
    {}


    \maketitle
%

\section{Introduction}
Protoplanetary discs are regulated by the interplay of a number of internal and external physical processes, which ultimately bring about the thousands of planets we observe.
Yet, despite their fundamental role, many questions remain unanswered as to how these processes operate together. One question relates to the identification and quantification of the driving angular momentum transport mechanisms that are active within the disc \citepads{2023ASPC..534..539M}. A second question concerns the role of the external environment in affecting the discs evolution, possibly replenishing them with material \citepads{2024A&A...691A.169W}, but ultimately driving their evaporation and dispersal \citepads{2022EPJP..137.1132W}. A third question relates to the response of the solid component, which eventually builds planetary cores, to both these internal and external processes \citepads{2023ASPC..534..717D}.

Concerning internal stresses, angular momentum transport in accretion discs was classically attributed to an effective turbulent viscosity. Given our ignorance on many of the physical details, a convenient way to model this process is via a dimensionless parameter $\alphaturb$ \citepads{1973A&A....24..337S}, which simply represents the ratio of radial stresses to pressure. A relatively large $\alphaturb\simeq 10^{-3} - 10^{-2}$ is needed to match observed accretion rates (\citealtads{1998ApJ...495..385H}, \citeyearads{2016ARA&A..54..135H}, \citealtads{2000prpl.conf..377C}, \citealtads{2007MNRAS.376.1740K}, \citealtads{2019A&A...631L...2M}).
Observationally, the turbulent $\alphaturb$ can be constrained between $10^{-4}$ and a few $10^{-2}$ (\citealtads{2016ApJ...816...25P}, \citealtads{2017ApJ...837..163R}, \citealtads{2018ApJ...869L..46D}, \citealtads{2017ApJ...843..150F}, \citeyearads{2018ApJ...856..117F}, \citealtads{2022ApJ...930...11V}; see \citealtads{2023NewAR..9601674R} for a review). 
Theoretical results show that $\alphaturb$ can be maintained above $\simeq 10^{-4}$ by various hydrodynamical or gravitational instabilities 
(e.g.\ \citealtads{1998MNRAS.294..399U}, \citealtads{1999ApJ...513..805L}, \citealtads{2001ApJ...553..174G}, \citealtads{2014A&A...572A..77S}, \citealtads{2015ApJ...804...62R}, \citealtads{2016ARA&A..54..271K}, \citealtads{2016MNRAS.455.2608L}, \citealtads{2021ApJ...915..130P}, \citealtads{2020ApJ...897..155F}; see \citealtads{2023ASPC..534..465L} for a recent review), while magneto-hydrodynamical (MHD) effects (such as the magneto-rotational instability, or MRI; \citealtads{1959JETP....9..995V}, \citealtads{1991ApJ...376..214B}) can drive $\alphaturb$ up to $10^{-2}$ in magnetically-active regions of the disc. 
MHD disc winds (\citealtads{1982MNRAS.199..883B}, \citealtads{1993ApJ...410..218W}, \citealtads{1997A&A...319..340F}, \citealtads{2009ApJ...691L..49S}) have been invoked as an alternative mechanism for angular momentum transport which has been recently gaining traction (\citealtads{2014prpl.conf..411T}, \citealtads{2016A&A...596A..74S}, \citealtads{2016ApJ...821...80B}, \citealtads{2022MNRAS.512.2290T}, \citeyearads{2022MNRAS.512L..74T}, \citealtads{2023ASPC..534..567P}). 
Whether discs are mainly viscously or MHD-wind driven is a topic of active research and observationally unresolved (e.g.\ \citealtads{2022MNRAS.514.1088Z}, \citealtads{2023MNRAS.524.3948A}, \citealtads{2023ApJ...954L..13S}, \citealtads{2024A&A...692A..93Z}, \citealtads{2025ApJ...989....7T}, \citealtads{2025arXiv251102811W}, and recent review by \citealtads{2023ASPC..534..539M}).\\

In fact, \citeads{2024MNRAS.527.7588C} recently showed that external processes can obfuscate the distinction between viscously-driven and MHD-disc-wind-driven angular momentum transport inside the disc, highlighting the importance of the interplay between internal stresses and the environment.
The dominant environmental influence is the irradiation from external EUV/FUV photons (e.g.\ \citealtads{1998ApJ...499..758J}, \citealtads{2004ApJ...611..360A}; see review by \citealtads{2022EPJP..137.1132W}). When a disc is subject to intense external radiation from other nearby massive stars, the heat produced in the outer layers can drive a thermal wind that allows material to escape the central star's potential well. This process removes mass (typically, FUV irradiation dictates the mass loss rate), truncates the disc, and thereby strongly affects disc evolution.\\

Despite its importance, external photoevaporation of protoplanetary discs is a poorly constrained process. The count of star forming regions (SFRs) with few stars (say, $\lesssim$ few hundred) is high, so these SFRs are commonly found close to Earth, are easily observed, and have been thoroughly covered by surveys (e.g.\ \citealtads{2013ApJ...771..129A}, \citeyearads{2018ApJ...869L..41A}, \citealtads{2016ApJ...828...46A}, \citeyearads{2018ApJ...859...21A}, \citealtads{2021MNRAS.506.5117T}, \citealtads{2016ApJ...831..125P}, \citealtads{2016ApJ...827..142B}, \citealtads{2018ApJ...869...17L}, \citealtads{2019MNRAS.482..698C}, \citealtads{2021A&A...645A.145G}, \citealtads{2025ApJ...989....1Z}). 
These regions have initial mass functions (IMFs) skewed towards low-mass K and M dwarfs where more massive stars (O/B) are less frequent. This makes them especially suitable for high-resolution campaigns (e.g.\ with the Atacama Large Millimeter Array or ALMA), since their proximity ($\lesssim 300\,\mathrm{pc}$) and relative isolation minimises contamination and extinction, and requires shorter observation times. On the other hand, they paint a rather biased picture of the environments in which protoplanetary discs evolve. As they host few massive stars, they are not subject to high external FUV fluxes, which are thus limited to $\sim 10\, \Gnaught$ (\citealtads{2018MNRAS.478.2700W}, \citealtads{2025ApJ...989....8A}, \citeyearads{2025A&A...695A..74A}; here, $\Gnaught$ is the Habing unit commonly used to measure FUV fluxes, \citealtads{1968BAN....19..421H}). However, SFRs with high number of stars (e.g.,\ Orion, Carina, or Cygnus OB2) host within them such a large number of them that a given star taken at random across all SFRs is more likely to have been born with a massive star nearby (e.g.\ \citealtads{2008ApJ...675.1361F}).
In other words, the typical protoplanetary disc observed in surveys so far is not representative of the planet-hosting star in the galaxy in terms of its neighbouring environment, where the FUV flux may exceed $100 - 1000\, \Gnaught$. \\

On the theoretical side, previous works have modelled the response of the gas to FUV fields \citepads{2007MNRAS.376.1350C}, including its dependence on the internal stresses \citepads{2024MNRAS.527.7588C}, others have investigated what happens to dust under external irradiation (e.g.,\ \citealtads{2020MNRAS.492.1279S}), or how theoretical predictions for purely viscous dusty discs within mild FUV environments relate to observations \citepads{2025ApJ...989....8A}. However, no study has systematically described the evolution of both the gas and the dust in viscously- vs. MHD-driven discs that are subject to external irradiation. This is the main aim of this paper.
The specific question we tackle is under which conditions the dust component can survive long enough so that there is enough time for planet formation to occur. To answer this question, we perform 1D simulations of the dust and gas components of discs with various internal stresses and different external FUV fluxes, and extract disc lifetimes that relate to both the gas and the dust.\\

The rest of the paper is organised as follows. In Section \ref{sec:DiscModel} we describe the disc model and the parameter space investigated in this work. In Section \ref{sec:Results} we present the main results for the evolution of gaseous and solid components. In Section \ref{sec:Discussion} we discuss their relevance in the context of our current understanding of disc evolution, disc observations, and the assembly of planetary systems, and we discuss the limitations of our modelling. Finally, we conclude in Section \ref{sec:Conclusions}.

\section{Disc model}\label{sec:DiscModel}
We begin by describing the underlying disc evolution model, for both the gas and the dust. We envision a protoplanetary disc around a central star of mass $\mStar$, assuming axi-symmetry and ignoring the vertical component for simplicity, making our model 1D in radius $r$. We treat the three main processes mentioned above, namely angular momentum transport mechanisms, external environment (specifically, external FUV irradiation) and dust evolution, in subsections \ref{subsec:AngMomTransp}, \ref{subsec:ExtPhotEvap} and \ref{subsec:Dust} respectively. The numerical setup and parameter space explored in this paper are described in subsection \ref{subsec:Code.ParamSpace}.

\subsection{Surface density evolution and angular momentum transport}\label{subsec:AngMomTransp}
To describe the surface density and angular momentum evolution of the disc, we use the hybrid viscous and MHD-disc-wind model of \citeads{2022MNRAS.512.2290T}, with the addition of a mass-removal term due to external photoevaporative winds (described in more detail in the next subsection). We briefly present the main equations here, and refer the reader to the original paper for a more in-depth presentation. The driving equation is the continuity equation for the gas surface density $\Sigma_\gas$ with sinks due to both MHD-wind-driven mass loss and external photoevaporation,
\begin{equation}\label{eq:basicMHDMaster.cylindrical.intz.parametrised}
\begin{split}
    \frac{\partial \Sigma_\gas}{\partial t} & - \frac{3}{r}\frac{\partial }{\partial r}\left[\frac{1}{r\Omega_\Kepl} \frac{\partial}{\partial r}\left[\alphaturb c_\sound^2 r^2 \Sigma_\gas \right]
    \right] \\
    & - \frac{3}{2r}\frac{\partial }{\partial r}\left[\frac{1}{\Omega_\Kepl} \alpha_\mathrm{DW} c_\sound^2 \Sigma_\gas \right]
    + \frac{3}{4} \frac{\alpha_\mathrm{DW} c_\sound^2 \Sigma_\gas}{(\lambda-1)\Omega_\Kepl r^2} \\
    & 
    + \dot{\Sigma}_{\gas,\extphot}(r) = 0.
\end{split}
\end{equation}
The orbital frequency is $\Omega_\Kepl = \sqrt{\GravC \mStar/r^3}$ (where $\GravC$ is the gravitational constant and $\mStar$ the star mass), such that the reference unperturbed azimuthal velocity $v_{\gas,\phi,0} = v_\Kepl = r \Omega_\Kepl$; the radial velocity of the gas is $v_{\gas,r}$ (in other words, its perturbation from the reference unperturbed value of 0). The isothermal sound speed $c_\sound$ is linked to surface density $\Sigma_\gas$ and (vertically-integrated) pressure $P$ by $P = \Sigma_\gas c_\sound^2$, to the aspect ratio $h=H/r$ by $h = c_\sound / v_\Kepl$ (where $H$ is the disc's vertical scale-height), and to the temperature $T$ by $c_\sound^2 = \frac{k_\mathrm{b} T}{\mu m_\mathrm{p}}$ (where $k_\mathrm{b}$ is the Boltzmann constant and $\mu$ is the mean molecular weight of the gas in units of the proton mass $m_\mathrm{p}$). 

We now turn to the parameters $\alphaturb$ and $\alpha_\mathrm{DW}$ that appear in eq.\ \eqref{eq:basicMHDMaster.cylindrical.intz.parametrised}. Following the approach of \citeads{1973A&A....24..337S}, these quantities conveniently parametrise the (time-averaged) radial stress tensor (describing the radial transport of angular momentum) and the vertical stress tensor (describing the extraction of angular momentum by the MHD wind) respectively, including both Reynolds and Maxwell stresses\footnote{
These $\alpha$ parameters effectively measure stress-to-pressure ratios, with the explicit definitions \citepads{2022MNRAS.512.2290T}
\begin{align*}
    \label{eq:alphaturb.param__alphaDW.param}
    \alphaturb := \frac{2}{3} \frac{\int_{-H_\wind}^{+H_\wind} T_{r \phi}\,\mathrm{d} z}{\Sigma_\gas c_\sound^2},
    \quad
    \alphaDW := \frac{4}{3} \frac{r \left[ T_{z \phi}\right]_{-H_\wind}^{+H_\wind}}{\Sigma_\gas c_\sound^2},
\end{align*}
where $T_{r \phi}$ and $T_{z \phi}$ are the components of the (time-averaged) stress tensor associated to the radial and vertical transport of angular momentum, and $H_\wind$ is the vertical scale height where the MHD wind is launched.
}.
Following \citeads{2022MNRAS.512.2290T}, we assume constant $\alphaturb$ and $\alphaDW$ for simplicity throughout. We then define a total $\alpha$ parameter
\begin{equation}\label{eq:def.alphatot}
    \alpha_\mathrm{tot}: = \alphaturb+\alphaDW;
\end{equation}
although practically convenient (\citealtads{2022MNRAS.512.2290T}, see also e.g.\ \citealtads{2022MNRAS.514.1088Z}), we note that no actual physical meaning should be attached to this quantity, as it sums $\alpha$ parameters arising from two processes of fundamentally different nature (diffusion and advection). We finally introduce the relative strength between the $\alpha$ parameters associated to the radial transport and vertical extraction of angular momentum \citepads{2022MNRAS.512.2290T}
\begin{equation}\label{eq:def.psiDW}
    \psiDW: = \frac{\alphaDW}{\alphaturb}.
\end{equation}
Thus, $\psiDW = 0$ represents a purely viscous disc, while $\psiDW\gg1$ represents a purely MHD-disc-wind evolution model.

As in \citeads{1973A&A....24..337S}, one can parametrise $\nu_\mathrm{SS} \Omega_\Kepl = \alphaturb c_\sound^2$ and $\nu_\mathrm{DW} \Omega_\Kepl = \alphaDW c_\sound^2$.
Thus, $\nu_\mathrm{SS}$ plays the usual role of the effective momentum diffusivity arising from radial stresses. We make use of the locally isothermal approximation, and write $c_\sound^2 \propto T \propto r^{-\betaT}$ with a constant $\betaT$.
We follow \citeads{2020MNRAS.492.1279S} and use a fiducial value of $\betaT=1/2$, so that $\nu_\mathrm{SS} \propto r^{\gamma_\mathrm{visc}}$ with $\gamma_\mathrm{visc} = 3/2 - \betaT = 1$. This profile is appropriate for the standard passive irradiated disc regime, where stellar irradiation dominates the dust temperature (\citealtads{1987ApJ...323..714K}, \citealtads{1997ApJ...490..368C}), and consistent with observational considerations (\citealtads{1998ApJ...495..385H}, \citealtads{2005ApJ...631.1134A}, \citealtads{2016ARA&A..54..135H}).

The last parameter in eq.\ \eqref{eq:basicMHDMaster.cylindrical.intz.parametrised} is the magnetic lever arm parameter $\lambda = \frac{h_\mathrm{DW}(r)}{r \Omega_\Kepl(r)}$, where $h_\mathrm{DW}(r)$ is the (specific) angular momentum carried away along the MHD-wind streamline anchored at $r$. One can show from energy conservation arguments that $\lambda>3/2$ and typical values from observations are consistent with values between 1.6 and 5 \citepads{2022MNRAS.512.2290T}. We take a reference fiducial value of $3$.
The second-to-last term in eq.\ \eqref{eq:basicMHDMaster.cylindrical.intz.parametrised} represents an MHD wind-driven mass-loss term (a sink),
\begin{equation}\label{eq:dotSigmagDW}
    \dot{\Sigma}_{\gas,\wind} := \frac{3}{4} \frac{\alphaDW c_\sound^2 \Sigma_\gas}{(\lambda-1)\Omega_\Kepl r^2}.
\end{equation}
Introducing the MHD disc wind velocity
\begin{equation}\label{eq:v_DW.def}
    v_\mathrm{DW} := \frac{3}{2} \frac{1}{r} \frac{\alphaDW c_\sound^2}{\Omega_\Kepl},
\end{equation}
we have $\dot{\Sigma}_{\gas,\wind} \equiv \frac{v_\mathrm{DW} \Sigma_\gas}{2(\lambda-1) r}$. 
The total (radial) local mass flux reads
\begin{equation}\label{eq:basicMHDAccrRate.parametrised.vDW}
    \Sigma_\gas v_{\gas,r} = -\frac{3}{r^2 \Omega_\Kepl} \frac{\partial}{\partial r}\left[\alpha_\mathrm{SS} c_\sound^2 r^2 \Sigma_\gas\right] - \Sigma_\gas v_\mathrm{DW},
\end{equation}
the first term being the local mass flux driven by turbulence, and the second being the local mass flux driven by the MHD wind.

For the initial conditions, like \citeads{2020MNRAS.492.1279S} (see also e.g.\ \citealtads{2022MNRAS.514.1088Z}, \citealtads{2025ApJ...989....8A}), we use the self-similar solution of \citeads{1974MNRAS.168..603L} for the surface density:
\begin{equation}\label{eq:SigmaInit}
    \Sigma_\gas(r; t=0) = \Sigma_{\gas,0} \left(\frac{r}{\Rc}\right)^{-\gamma_\visc} \exp\left(-\frac{r}{\Rc}\right)^{2-\gamma_\visc}.
\end{equation}
The reference cut-off radius $\Rc$ sets the initial disc size, and $\gamma_\mathrm{visc} = 3/2 - \betaT$, which equals $1$ with our fiducial $\betaT = 1/2$.

\subsection{External photoevaporation}\label{subsec:ExtPhotEvap}
Our treatment of external photoevaporation follows closely that of \citeads{2020MNRAS.492.1279S} and \citeads{2025ApJ...989....8A}. We prescribe a photoevaporative mass-loss rate $\dot{\Sigma}_{\gas,\extphot}(r)$ in eq.\ \eqref{eq:basicMHDMaster.cylindrical.intz.parametrised} as follows.
We use the updated FRIEDv2 grid of \citeads{2023MNRAS.526.4315H} with a fiducial ratio between the Polycyclic Aromatic Hydrocarbon (PAH) abundance and the dust abundance of 1 (see Appendix \ref{apx:FRIEDv1.vs.v2} for a quantitative comparison with the original FRIED grid from \citealtads{2018MNRAS.481..452H}, used e.g.\ in \citealtads{2020MNRAS.492.1279S}). Given a fixed $\mStar$ and an FUV flux, we obtain a local mass-loss rate $\dot{\tilde{M}}_{\gas,\extphot,i}$ at each radius $r_i$ (where $r_i$ are the radial cell centres used in the code as described in Subsect.\ \ref{subsec:Code.ParamSpace}), using a bilinear interpolation of the FRIEDv2 grid over $r_i$ and $\Sigma_\gas(r_i)$. This $\dot{\tilde{M}}_{\gas,\extphot,i}$ is, however, not the actual local mass-loss rate that should be used. This is because, as shown by \citeads{2020MNRAS.492.1279S}, the photoevaporation rate first increases with $r_i$ in the weakly-bound regions that are optically thick in the FUV and then decrease with $r_i$ in the optically thin region. However, the flow in the outermost, optically thin disc region is actually set by the larger rates from the closer-in, optically thick disc regions. Thus, simply applying the local mass-loss rate from the FRIEDv2 grid at $r_i$, taken at face value, would be unphysical. This is because the local mass-loss rate in the optically thin region assumes that the region interior to it does not flow in the wind, which is in reality not the case. \citeads{2020MNRAS.492.1279S} showed that the actual mass-loss rate is set at (and slightly outside of) the radius located at the transition between the optically thick and optically thin regimes. This so-called truncation radius $R_\trunc$ (cfr.\ also \citealtads{2025ApJ...989....8A}) is the radial location which maximises the mass-loss rate $\dot{\tilde{M}}_{\gas,\extphot,i}$ (the one obtained from the FRIED grid assuming the radial cell centre $r_i$ to be the outer radius of the disc, \citealtads{2020MNRAS.492.1279S}). Mass removal by external FUV can thus be implemented inside-out, at radii $r > R_\trunc$ (i.e.,\ over the radial grid, at radii $r_i$ with $i > i_\trunc$). First, we obtain a total mass-loss rate given by
\begin{equation}
    \dot{M}_{\gas,\tot,\extphot} = \sum_{i > i_\trunc} \dot{\tilde{M}}_{\gas,\extphot,i} \frac{M_{\gas,i}}{M_\gas(r > R_\trunc)}.
\end{equation}
Then, this mass-loss rate is distributed at larger radii in proportion to the local mass. The effective local mass-loss rate is thus
\begin{equation}
    \dot{M}_{\gas,\extphot,i} = \dot{M}_{\gas,\mathrm{tot},\extphot} \frac{M_{\gas,i}}{M_\gas(r > R_\trunc)}.
\end{equation}
This yields the sink term in the surface density evolution equation (Eq.\ \eqref{eq:basicMHDMaster.cylindrical.intz.parametrised}) with $\dot{M}_{\gas,\extphot}(r) = 2\pi r\, \Delta r\, \dot{\Sigma}_{\gas,\extphot}(r)$.

\subsection{Dust component}\label{subsec:Dust}
Equation \eqref{eq:basicMHDMaster.cylindrical.intz.parametrised} describes the evolution of the gas.
When dealing with dust, we use the two-population model of \citeads{2012A&A...539A.148B} (see also \citealtads{2017MNRAS.469.3994B}, \citealtads{2020MNRAS.492.1279S}), which we briefly describe here. The dust component can either be in a small grain population (with fixed monomer size $a_0 = 0.1\,\mu\mathrm{m}$, which is also the initial size of the dust; cfr.\ \citealtads{2020MNRAS.492.1279S}) or in a large grain population with size $a_1$, set by growth and limited by fragmentation and radial drift (which are the dominant factors in limiting growth in the inner and outer portions of the disc, respectively, \citealtads{2010A&A...513A..79B}). The two species have surface densities $\Sigma_{\dust,k}$, $k=0,1$, and the dust mass fractions are denoted by $\epsilon_{\dust,k} = \Sigma_{\dust,k}/\Sigma_\tot$, where $\Sigma_\tot = \Sigma_\gas + \Sigma_{\dust,0} + \Sigma_{\dust,1}$ is the total gas and dust surface density. The total dust mass fraction is $\epsilon_\dust:=\epsilon_{\dust,0}+\epsilon_{\dust,1}$ and is initialised completely in the small grains population as $\epsilon_\dust(t=0) = \epsilon_{\dust,0}(t=0) = 10^{-2}$ \citepads{1978ApJ...224..132B}.

Dust evolution is driven by advection following the motion of the gas, radial drift due to drag from the gas, diffusion, and removal from external photoevaporation due to FUV irradiation. For each dust species the evolution is thus governed by (e.g.,\ \citealtads{2010A&A...513A..79B})
\begin{equation}\label{eq:basicDustMaster.cylindrical}
    \frac{\partial \Sigma_{\dust,k}}{\partial t} + \frac{1}{r} \frac{\partial}{\partial r} \left[r \left(\Sigma_{\dust,k} v_{\dust,k,r} -D_{\dust,k} \frac{\partial}{\partial r}\left[\frac{\Sigma_{\dust,k}}{\Sigma_\gas}\right] \Sigma_\gas\right)\right] = s_{\dust,k}.
\end{equation}
The dust radial velocity $v_{\dust,k,r}$ has a drag component and radial drift component (e.g.,\ \citealtads{1977MNRAS.180...57W}, \citealtads{2014prpl.conf..339T})
\begin{equation}
    v_{\dust,k,r} = \frac{1}{1+\St_k^2} v_{\gas,r} + \frac{2\St_k}{1+\St_k^2} v_P,
\end{equation}
where $v_{\gas,r}$ is the radial velocity of the gas (cfr.\ eq.\ \eqref{eq:basicMHDAccrRate.parametrised.vDW}), $v_P=-\eta v_\Kepl$ and $\eta=-\frac{1}{2} \left(\frac{H}{r}\right)^2 \frac{\partial\log P}{\partial\log r}$.
Moreover, $v_{\dust,k,r}$ strongly depends on the Stokes number $\St_k$, which in the Epstein drag regime can be written as \citepads{2012A&A...539A.148B}
\begin{equation}
    \St_k = \frac{\pi}{2} \frac{a_k \rho_\solid}{\Sigma_\gas},
\end{equation}
assuming spherical particles with radius $a_k$ and material density $\rho_\solid$, which we take to be $1\,\mathrm{g}\,\mathrm{cm}^{-3}$. 
The diffusion coefficient $D_{\dust,k} = \frac{D_\gas}{\Sc_k}$, where the Schmidt number $\Sc_k = 1+\St_k^2$ \citepads{2007Icar..192..588Y}, and we take $D_\gas = \nu_\gas = \nu_\mathrm{SS}$, simply the gas turbulent viscosity. Dust feedback onto the gas is included as in \citeads{2017MNRAS.469.3994B} and \citeads{2020MNRAS.492.1279S}, although it makes little difference due to small dust mass fractions.

For the sink term $s_{\dust,k}$ in eq.\ \eqref{eq:basicDustMaster.cylindrical}, we assume that dust is only removed by external photoevaporation (but not by the MHD disc wind, see below). To model this process, we follow \citeads{2016MNRAS.457.3593F} and assume that dust particles smaller than a size \begin{equation}
    a_\mathrm{ent} = \frac{1}{4\pi \mathcal{F}} \frac{v_\mathrm{th} \dot{M}_{\gas,\extphot}}{\GravC \mStar \rho_\solid}
\end{equation}
are entrained in the photoevaporative wind, as the drag force produced by the wind on a grain smaller than this size overcomes the gravitational force on the grain due to the central star. In the formula above, $\mathcal{F} = \frac{H}{\sqrt{r^2 + H^2}}$ is a geometric factor so that $4\pi \mathcal{F}$ is the fraction of the solid angle subtended by the disc outer edge, and $v_\mathrm{th}$ is the thermal velocity of the gas. With this, the entrained mass fraction $f_\mathrm{ent}$ can be worked out for a size distribution $n(a)\,\mathrm{d}a \propto a^{-p}\,\mathrm{d}a$ as \citepads{2020MNRAS.492.1279S}
\begin{equation}
    f_\mathrm{ent} = \min \left[1, \left(\frac{a_\mathrm{ent}}{a_1}\right)^{4-p}\right],
\end{equation}
and the dust mass removed (locally) by the photoevaporative wind is thus $\dot{M}_{\dust,\extphot} = f_\mathrm{ent} \frac{M_\dust}{M_\gas} \dot{M}_{\gas,\extphot}$ (all quantities evaluated at each grid cell).

We note that we assume that the MHD disc wind does not remove any dust, that is, there is no $\dot{\Sigma}_{\dust,k,\wind}$ contribution to $s_{\dust,k}$. The main reason for this is that the amount of dust removed by these winds represents only a small fraction of the total dust mass and including this effect would not significantly modify the picture (see also Appendix \ref{apx:DWDustEntr_RD}, and the discussion in sect.\ 6.2 of \citealtads{2022MNRAS.514.1088Z}).
%
\begin{center}%
\begin{table}[t!]
\centering
\begin{tabular}{ m{1.5cm} m{2.3cm} m{4.1cm} } 
 \hline
 Parameter							& Value(s)  						& Description \\ \hline \hline 
 $M_*\,[M_\odot]$					& $1$ 								& Stellar mass\\
 $R_*\,[R_\odot]$					& $2.5$ 							& Stellar radius \\
 $T_\mathrm{eff,*}\,[\mathrm{K}]$	& $4000$ 							& Stellar effective temperature \\
 $M_{\disc,0}\,[M_\mathrm{Jup}]$	& $100$ 							& Initial disc mass \\
 $h_0$								& $0.033$ 							& Disc aspect ratio at $1\,\mathrm{AU}$\\
 $\betaf$							& $1/4$ 							& Flaring index \\
 \hline 
 $\alpha_\tot$						& $\{10^{-4}, 10^{-3}, 10^{-2}\}$	& $\alphaturb+\alphaDW$ (Eq.\ \eqref{eq:def.alphatot}) \\
 $\psiDW$							& $\{0, 0.1, 1, 10, 10^{10}\}$		& $\alphaDW/\alphaturb$ (Eq.\ \eqref{eq:def.psiDW})\\
 $\Rc\,[\mathrm{AU}]$				& $\{10, 30, 100\}$					& Initial disc cut-off radius\\
 $G\,[\Gnaught]$					& $\{0, 10, 100, 1000\}$			& External FUV flux \\
 \hline
\end{tabular}
\caption{Parameters of our simulations.}
\label{tbl:parameters}
\end{table}
\end{center}%
\subsection{Numerical scheme and parameter space}\label{subsec:Code.ParamSpace}
We solve the gas and dust evolution using a version of the \lstinline{DiscEvolution} code originally described in \citealtads{2017MNRAS.469.3994B} (cfr.\ also Appendix \ref{apx:CodeEqs}) while the external photoevaporation routines follow \citeads{2020MNRAS.492.1279S} (cfr.\ also \citealtads{2025ApJ...989....8A}). We use a radial grid from $R_\mathrm{min}=0.1\,\mathrm{AU}$ to a fiducial $R_\mathrm{max}=500\,\mathrm{AU}$, with a resolution of $N_r = 500$ and a natural spacing $\propto r^{1/2}$ (similar to \citealtads{2025ApJ...989....8A}, with a slightly higher radial resolution). When $\alphaturb$ is above $0.5 \times 10^{-3}$ and the disc is already initially large ($\Rc \geq 100\,\mathrm{AU}$), we extend the disc up to $R_\mathrm{max}=1000\,\mathrm{AU}$ and modify $N_r = 700$ accordingly. This is to avoid the outer edge of the disc from reaching too close to the grid outer boundary. This also allows us to safely use a zero-torque boundary condition at the outer edge, while we use an inner boundary condition that keeps a constant flux, appropriate for power law profiles (note that the inner disc quickly reaches a steady state, \citealtads{2022MNRAS.512.2290T}). We experimented with different radial extents, resolutions and spacings and found consistent results.\\

Our fiducial setup consists of a disc around a $\mStar=1\,\mSun$ star (with a reference $R_* = 2.5\,R_\odot$ and $T_\mathrm{eff,*} = 4000\,\mathrm{K}$) with initial disc mass $M_{\disc,0} = 100\,M_\mathrm{Jup}$ (i.e.,\ $M_{\disc,0} \simeq 10^{-1}\,\mStar$), which sets $\Sigma_{\gas,0}$ in equation \eqref{eq:SigmaInit}.
We also fix a reference aspect ratio $h_0 = 0.033$ at $1\, \mathrm{AU}$, while, given our locally isothermal assumption, $h(r)\propto r^\betaf$ with $\betaf = (-\betaT+1)/2$. Thus, $h(r)\propto r^{1/4}$ for our fiducial $\betaT = 1/2$. This gives a reference temperature of $\simeq280\,\mathrm{K}$ at $1\, \mathrm{AU}$ for a solar-mass star.
We explore a grid of parameters $\alpha_\tot \in \{10^{-4}, 10^{-3}, 10^{-2}\}$, $\psiDW \in \{0, 0.1, 1, 10, 10^{10}\}$ (where $\psiDW = 10^{10}$ represents the purely MHD-wind disc), a reference magnetic lever arm parameter $\lambda=3$, external FUV fluxes $G/\Gnaught \in \{0, 10, 100, 1000\}$ (with a nominal PAH fraction $f_\mathrm{PAH} = 1)$, and initial disc cut-off radii $\Rc/\mathrm{AU} \in \{10, 30, 100\}$. Table \ref{tbl:parameters} presents the full set of parameters. In a companion paper, we explore the dependence of our results on underlying model parameters such as the initial disc mass, the stellar mass, the PAH fraction and the magnetic lever arm parameter (Pichierri et al.\ \emph{in prep.}).

\begin{figure*}[h]
    \centering
    \includegraphics[width=1\linewidth]{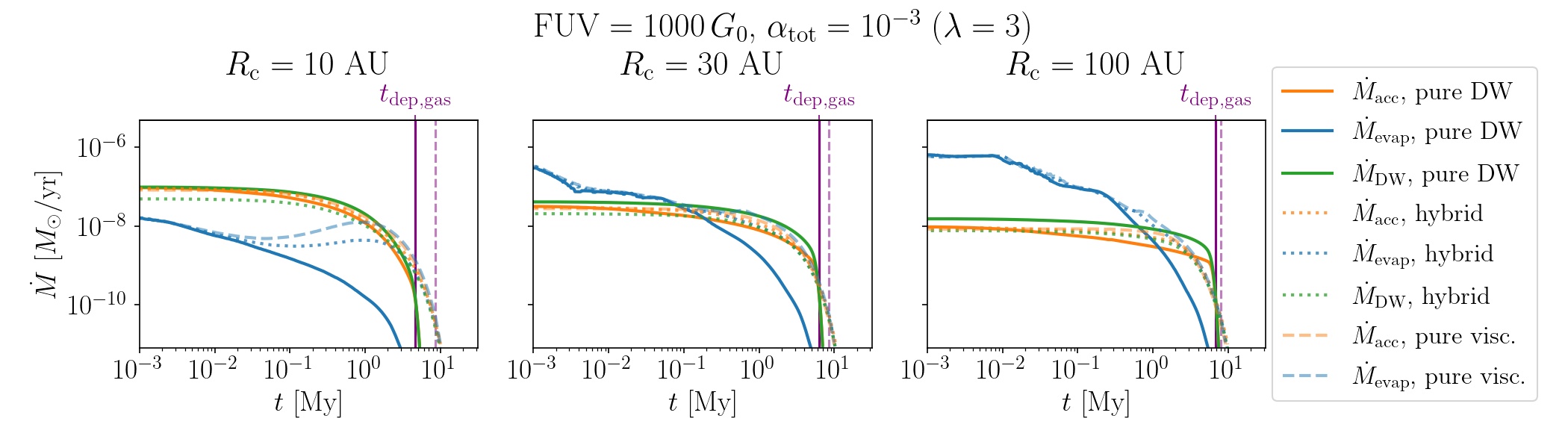}
    \includegraphics[width=1\linewidth]{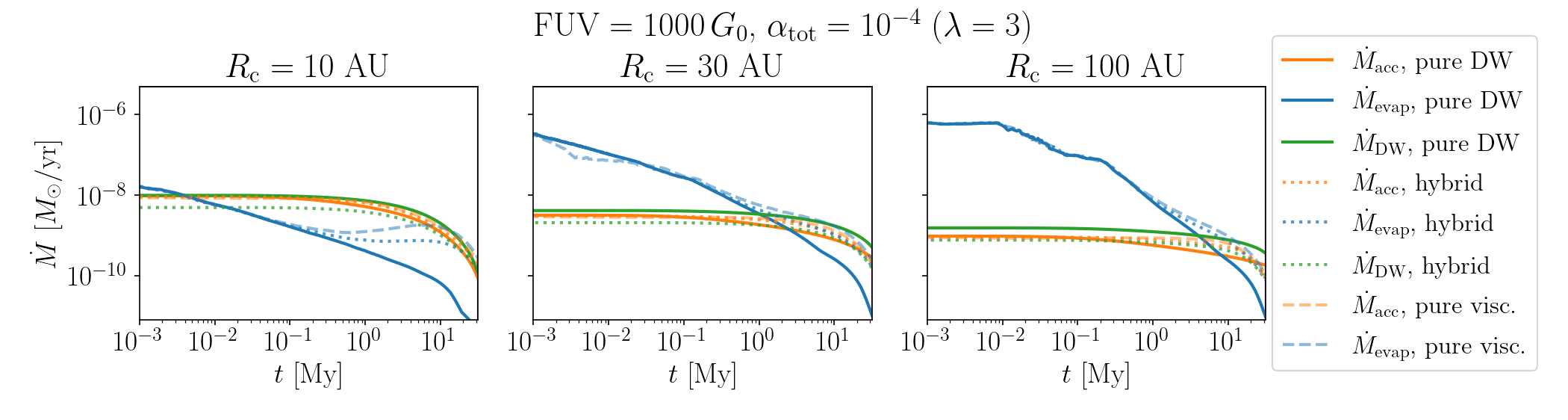}
    \caption[]{The mass accretion rates, photoevaporative mass loss rates and MHD-wind-driven mass loss rates for discs of different initial sizes and with different values of $\psiDW$, namely $\psiDW=0$ (purely viscous disc, dotted lines), $\psiDW=1$ (hybrid disc, dashed lines), and $\psiDW\gg1$ (purely MHD-wind disc, continuous lines). For explanatory purposes we consider here high irradiation, $1000\,\Gnaught$, and low-to-moderate total $\alpha_\tot = 10^{-4} - 10^{-3}$. The purple vertical dashed lines represent the gas disc lifetime diagnostic $\tMg$ (cfr.\ subsect.\ \ref{subsec:Diagnostics})} 
    \label{fig:dotMs_FUV=1000_lam=3}
\end{figure*}

\section{Results}\label{sec:Results}
\begin{figure*}[h]
\centering
\includegraphics[width=1.0\textwidth]{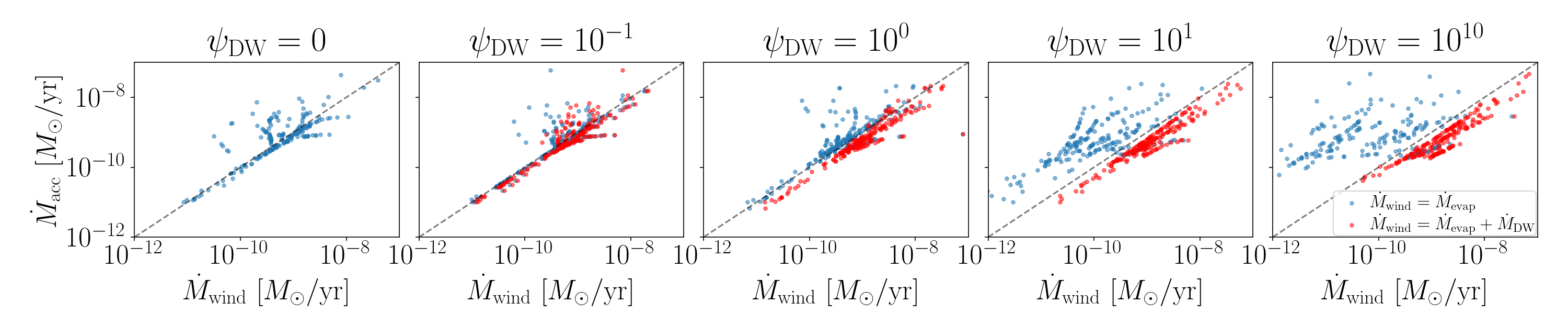}%
\caption[]{The mass-accretion rate vs.\ the mass-loss rate in the wind (either the photoevaporative wind alone -- blue dots --, or combined with the MHD wind -- red dots), obtained from mock observations at random times, selected randomly from a uniform distribution in the interval $[0\,\mathrm{My},31.6\,\mathrm{My}]$. Each panel shows the outcome of these experiments for different fixed underlying $\psiDW = \alphaDW/\alphaturb$, while the initial disc sizes, total $\alpha_\mathrm{tot}$, and (finite) FUV fluxes are drawn randomly (values from Table \ref{tbl:parameters}).}
\label{fig:mock_pop_synth}
\end{figure*}

\subsection{Gas evolution}\label{subsec:GasEvo}
For the gas component alone, the coupling of purely viscously-evolving discs (VE discs for short) with external photoevaporation has been modelled for the first time by \citeads{2007MNRAS.376.1350C}, who showed that these discs follow a predictable evolution. As long as the accretion rate onto the star is larger than the mass-loss rate at the photoevaporative edge, the disc expands outwards and self-adjusts to a state where the outward viscous flow at the outer edge matches $\dot{M}_\mathrm{evap}$. When $\dot{M}_\mathrm{acc}$ is no longer larger than the photoevaporative rate, the disc outer edge moves inward on a timescale comparable to the local viscous time-scale. Thus, VE discs experience a long-lived phase where the accretion rate onto the star matches the photoevaporative rate; they also lose memory of the initial disc cutoff radius $\Rc$ \citepads{2020MNRAS.492.1279S}. The evolution is expected to be different in the MHD-wind case (DW discs for short): these discs do not expand outwards (e.g.\ review by \citealtads{2023ASPC..534..539M}), there is no restoring mechanism to reach a self-adjustment between the inner and outer regions, and the MHD-disc-wind mass removal mechanism is decoupled from external photoevaporation. Because there is no natural equilibrium state, a pure DW disc is then not independent of initial radius $\Rc$ at later times. These trends are shown in Figure \ref{fig:dotMs_FUV=1000_lam=3} for a highly irradiated disc, $1000\,\Gnaught$, and with moderate total $\alpha_\tot = 10^{-3}$ (top row), as a function of different initial disc radii.

Another consequence of the qualitatively different behaviour in VE vs.\ DW discs is that, for a purely DW disc, the long-lived phase of disc evolution has $\dot{M}_\mathrm{acc} > \dot{M}_\mathrm{evap}$. This is because either $\dot{M}_\mathrm{acc} > \dot{M}_\mathrm{evap}$ from the start (for compact discs), or (for larger disc) $\dot{M}_\mathrm{evap}$ quickly drops within less than 1 Myr, as material is not replenished at larger radial separations. Therefore, observing discs at random will preferentially reveal them at this evolutionary stage. Instead, sampling VE discs will typically reveal a $\dot{M}_\mathrm{acc} - \dot{M}_\mathrm{evap}$ correlation. This explains the recent results from \citeads{2025arXiv251101972W}, which propose distinguishing between the two angular momentum transport mechanisms by looking at accretion and evaporative mass loss rates (they specifically focused on Cygnus OB2, with FUV fluxes of order $10^2 - 10^4$; Fig.\ \ref{fig:dotMs_FUV=1000_lam=3} uses fluxes of $10^3\,\Gnaught$ for comparison). In particular, in the viscous case we expect an almost 1:1 relation between the two rates, while this is not the case for the (pure) MHD wind model.
We note, however, that this distinction is almost completely lost in the hybrid model ($\psiDW = 1$), where the little viscosity $\alphaturb$ is sufficient to recover the viscous spreading effects. To drive this point further, we perform mock-population synthesis experiments where we sample discs with a fixed $\psiDW$, across (finite) FUV fluxes, $\alpha_\mathrm{tot}$, and $\Rc$ randomly chosen across their respective ranges, performing ``observations'' at random times with a uniform distribution $\mathcal{U}_{[0\,\mathrm{My},31.6\,\mathrm{My}]}$. We plot in Figure \ref{fig:mock_pop_synth} the observed $\dot{M}_\mathrm{acc}$ vs.\ $\dot{M}_\mathrm{wind}$, defined as either $\dot{M}_\mathrm{wind} = \dot{M}_\mathrm{evap}$ or $\dot{M}_\mathrm{wind} = \dot{M}_\mathrm{evap} + \dot{M}_\wind$. Although very simplistic, this experiment shows how the $\dot{M}_\mathrm{acc} - \dot{M}_\mathrm{evap}$ correlation disappears with higher and higher $\psiDW$.

We should keep in mind that this analysis does not include the role of internal photoevaporative winds. Moreover, \citeads{2025arXiv251102811W} recently found that it may be difficult distinguish between thermally-driven and magnetically-driven mass loss, which may obfuscate the distinction even further.

\subsection{Dust evolution}\label{subsec:DustEvo}
\begin{figure*}[h]
\centering
\includegraphics[width=1.025 \textwidth]{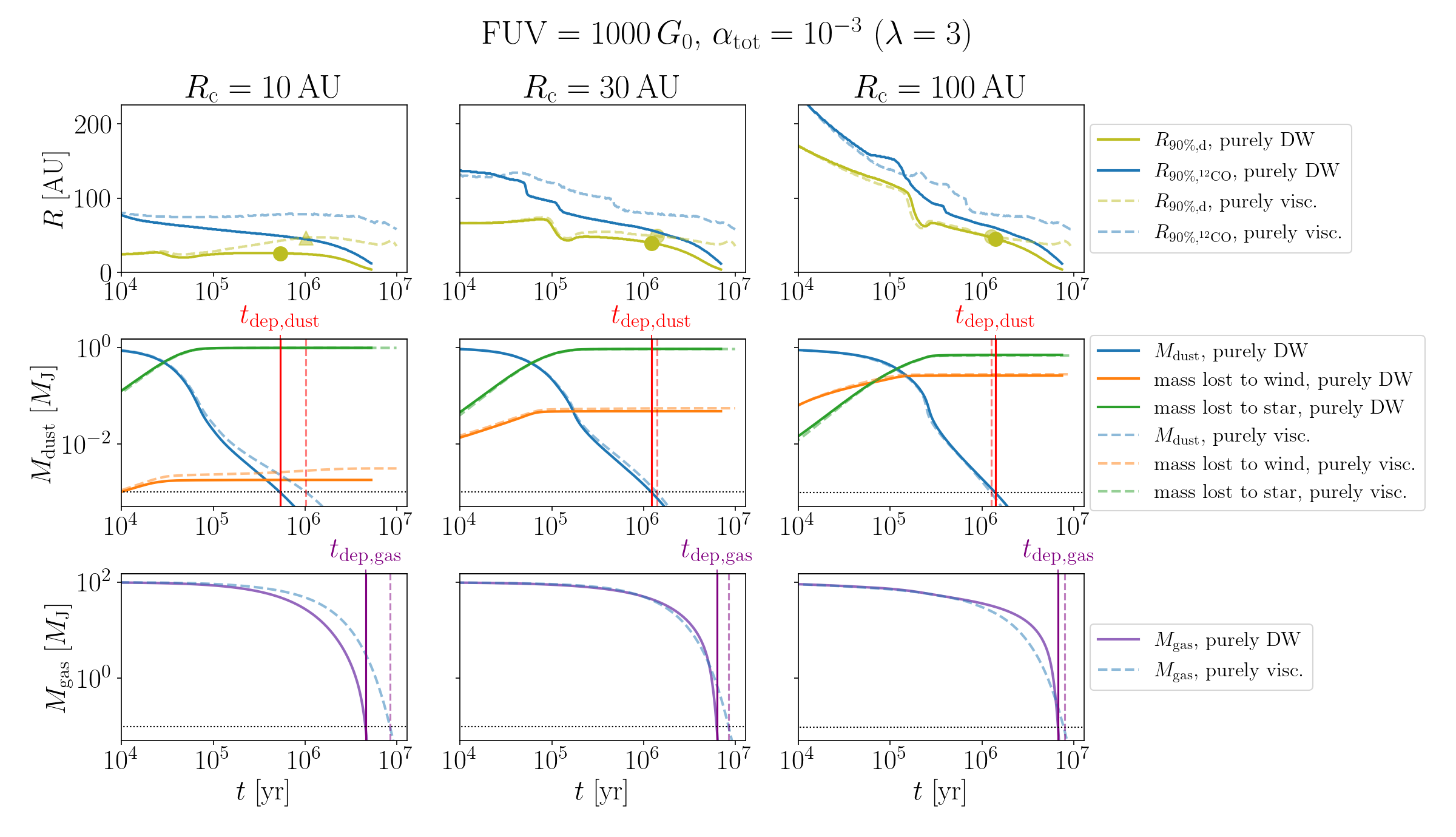}
\caption[]{
Typical evolution of representative dust and gas quantities for either fully MHD-driven discs (continuous lines) or fully viscous discs (dashed lines), subjected to external photoevaporation, as a function of initial disc size $\Rc$ (different columns). \emph{Top row}: dust radii $R_{90\%,\dust}$ enclosing 90\% of the dust mass, and $R_{90\%,{}^{12}\mathrm{CO}}$, the radius which encloses 90\% of the ${{}^{12}\mathrm{CO}}$ emission, calculated following {\citeads{2023ApJ...954...41T}}. Upward triangles label the cases where the dust radii initially expand, while circles indicate those that do not, due to external irradiation. \emph{Middle row}: time evolution of the total dust mass, mass accreted onto the star, and mass lost to wind. The vertical red lines denote the times when the dust mass reaches fraction of the initial mass (set to 1/1000). \emph{Bottom row}: time evolution of the gas mass. Similarly to the dust, the vertical purple lines denote the times when the gas mass reaches 1/1000 of the initial mass. The underlying total $\alpha_\mathrm{tot}=10^{-3}$ and the FUV flux is set to $1000\,\Gnaught$.
}
\label{fig:Evo__FUV=1000_ALPHA=1e-03_lam=3}
\end{figure*}

\begin{figure*}[h]
\centering
\includegraphics[width=0.95 \textwidth]{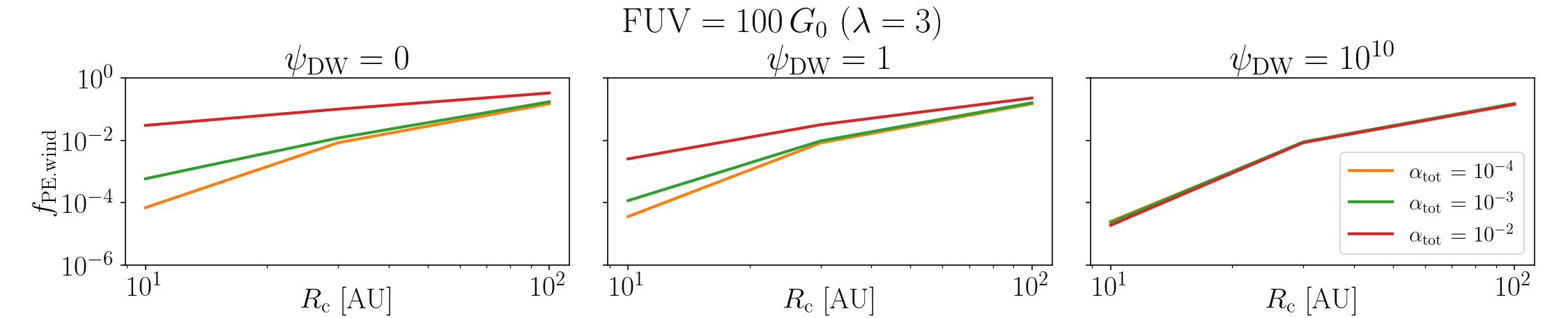}%
\caption[]{
Fraction of dust mass lost to wind as a function of the initial disc size $\Rc$, for different $\psiDW=\alphaDW/\alphaturb$ and $\alpha_\mathrm{tot}=\alphaDW + \alphaturb$.
}
\label{fig:f_wind.w.r.t.Rc}
\end{figure*}
Dust evolution, in both the VE and DW cases, is driven by growth, radial drift and removal from external photoevaporation, and these processes compete with one another in making dust lifetimes longer or shorter.
Dust that is well-coupled to the gas can follow its outward expansion in the viscous case.
Similarly, dust diffusion is active when $\psiDW\ll1$ (VE discs) but not in the MHD-wind case (because $D_{\dust} = \nu_\mathrm{SS}/{\Sc}$, cfr.\ subsect.\ \ref{subsec:Dust}).
Since the photoevaporative mass-loss rate increases with radius in the optically-thick regime \citepads{2020MNRAS.492.1279S}, external irradiation would cause DW discs to show longer lifetimes than their purely viscous counterparts.
Instead, drift causes dust to move inward and accrete onto the star. When $\psiDW\gg1$ (DW discs), because dust does not diffuse outward (like it does in the viscous case) it also has less of a radial displacement to cover; the relative efficiency of drift and gas drag will depend on the Stokes number. The fact that DW discs do not diffuse outwards also means that their surface density profiles in the outer disc will be steeper as they are carved out by external FUV irradiation, so that inward drift can be more efficient the outer regions depending on the disc's initial size.
Just like for the gas, an important difference is in the discs' ability to expand via diffusion, which is fundamentally different for VE and DW discs (see \citealtads{2022MNRAS.514.1088Z}). But, just like the gas \citepads{2024MNRAS.527.7588C}, this difference can be strongly suppressed by high external FUV fluxes, which can obfuscate the distinction between the different angular momentum transport mechanisms.

The top row of Figure \ref{fig:Evo__FUV=1000_ALPHA=1e-03_lam=3} shows the time evolution of $R_{90\%,\dust}$, defined as the radius that encloses 90\% of the dust\footnote{
For completeness, we add the radius $R_{90\%,{}^{12}\mathrm{CO}}$ which encloses 90\% of the ${{}^{12}\mathrm{CO}}$ emission. This is calculated as the location where column density is below a critical value, inhibiting self-shielding and causing photo-dissociation, $N\approx 10^{21.27 - 0.53 \log(L_*/L_\odot)} \times \left(\frac{M_\gas}{M_\odot}\right)^{0.3-0.08\log(L_*/L_\odot)}\,\mathrm{cm}^{-2}$ (\citealtads{2023ApJ...954...41T}; see also \citealtads{2025ApJ...989....8A}).
}, for an underlying FUV flux of $1000\,\Gnaught$, $\alpha_\mathrm{tot} = 10^{-3}$ and for various different $\Rc$ values. The total $\alpha_\mathrm{tot}$ is distributed either completely into $\alphaturb$ ($\psiDW = 0$, purely viscous case, dashed transparent lines) or completely into $\alphaDW$ ($\psiDW \gg 1$, purely DW case, continuous opaque lines). The biggest difference arises in the $\Rc = 10\,\mathrm{AU}$ case, which, given the high FUV flux, is the only one here where the purely viscously-evolving disc has a chance to expand. Across the top row of Figure \ref{fig:Evo__FUV=1000_ALPHA=1e-03_lam=3}, the ability of the dust disc to expand outwards is marked by an upward triangle, while we use a round marker for those setup where the radii do not significantly expand (for purely visual reasons, these symbols are positioned at the time $\tMd$ when the dust disc is 1/1000 of the initial mass, see also second row of Fig.\ \ref{fig:Evo__FUV=1000_ALPHA=1e-03_lam=3} and next subsection). 

Because of drift, dust evolution happens on a different timescale than gas evolution, as can be seen comparing the bottom two rows of Fig.\ \ref{fig:Evo__FUV=1000_ALPHA=1e-03_lam=3}, specifically the vertical red and purple lines which represent dust and gas depletion timescales (see also Appendix \ref{apx:DWDustEntr_RD}).
As we show in greater detail in the next subsection, dust evolution typically happens faster (although marginally so) for DW discs compared to VE discs (see e.g.\ Fig.\ \ref{fig:Evo__FUV=1000_ALPHA=1e-03_lam=3}, specifically the vertical red lines which measure the dust depletion timescales on the top two rows).
This result differs slightly from \citeads{2025arXiv251104410C}, which use a population synthesis approach to conclude that the median dust masses decrease faster for viscously evolving discs rather than MHD wind discs when exposed to external photoevaporation. This difference most likely arises from the fact that they do not explicitly solve for the evolution of the solid component as an additional species.

As expected, initially larger discs lose increasingly more dust to the photoevaporative (PE) wind, as shown in Figure \ref{fig:f_wind.w.r.t.Rc} as a function of the dust mass fraction lost to the PE wind $f_\mathrm{PE.wind}$, and in accordance with \citeads{2020MNRAS.492.1279S}. However, one difference of note is that the clear dependence on $\alpha$ for VE discs (see also Fig. \ref{fig:v1.vs.v2}, right panel) is less and less measurable for DW discs: the amount of dust mass lost mostly depends on $\alphaturb$ in VE discs because of viscous spreading, while in the MHD wind case it mostly depends on initial disc size. Also note that, while larger discs are subjected to greater dust mass removal from external photoevaporation, this does not lead to them having shorter dust lifetimes because more compact discs tend to lose their dust more quickly onto the star.

\subsection{Disc lifetime diagnostics}\label{subsec:Diagnostics}
\begin{figure*}[h]
\centering
\includegraphics[width=1.025\textwidth]{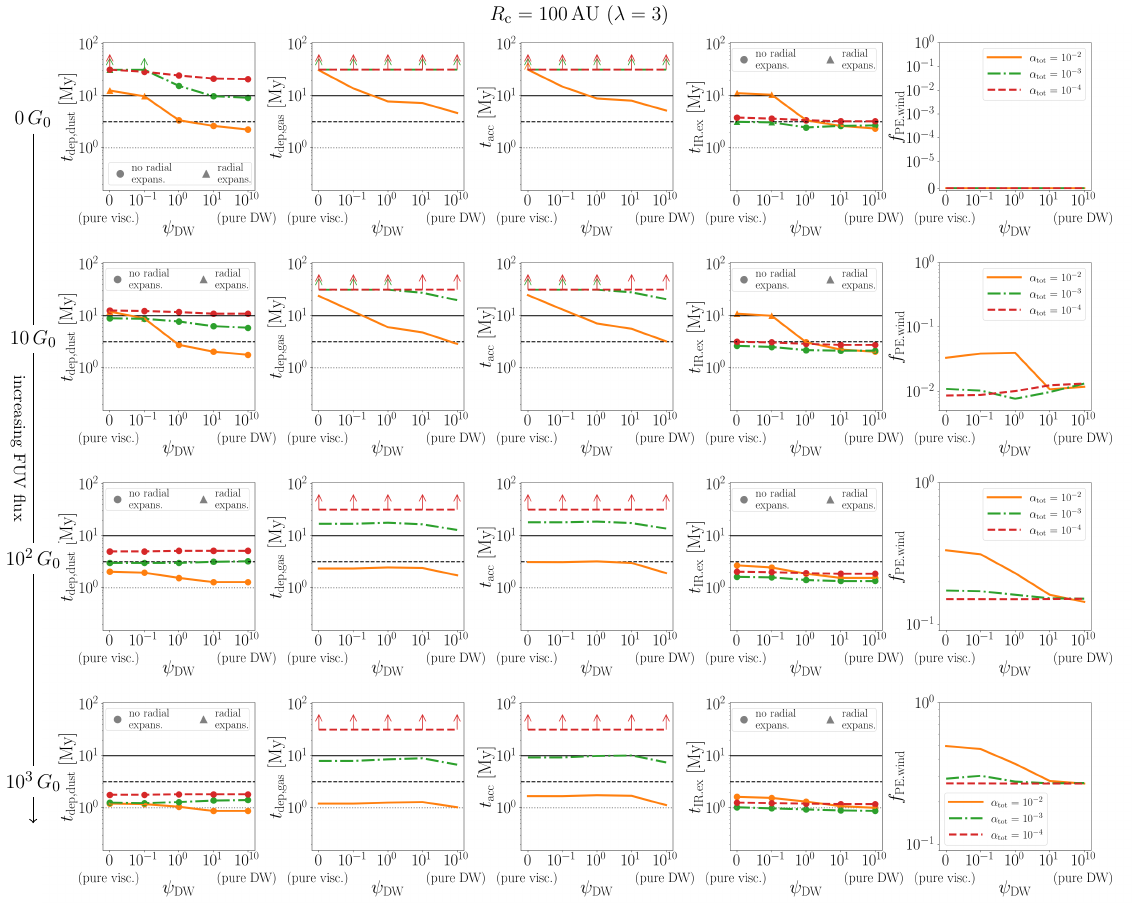}

\caption[]{
Disc diagnostics as a function of $\psiDW = \alphaDW/\alphaturb$ for different total $\alpha_\mathrm{tot} = \alphaDW + \alphaturb$ (different colours and line styles, see legends in the last column), and for different FUV fluxes over different rows (see labels on the left along the arrow). In all cases, the disc's initial size $\Rc = 100\,\mathrm{AU}$. Upward pointing arrows at $t=31.6\,\mathrm{My}$ show that the data points represent lower bounds.
\emph{Left column}: $\tMd$, the time when the total dust mass drops below 1/1000 of the initial dust mass. \emph{Second column}: $\tMg$, the time when the gas mass drops below 1/1000 of the initial gas mass. \emph{Third column}: $\tMacc$, the time at which the accretion rate onto the star drops below $10^{-11}\,\mSun/\mathrm{yr}$. \emph{Fourth column}: $\tIRex$, the time when the disc would be undetectable due to lack of IR excess; we estimate this as the time when the surface density of \emph{small} dust at $r=1\,\mathrm{AU}$ drops below $1/\kappa_0$, with a reference opacity of small dust grains $\kappa_0 = 1000\,\mathrm{cm}^2/\mathrm{g}$. \emph{Last column}: fraction of dust mass lost to wind (note the somewhat inconsistent vertical axis range due to values spanning different orders of magnitude across the different setups; see also Fig.\ \ref{fig:f_wind.w.r.t.Rc}).
As shown in the legend in the first and fourth columns (which relate to dust timescales), the upward triangles indicate runs where the dust disc radii initially expand due to significant diffusion, while circles indicate those setups which do not (cfr.\ top row of Fig.\ \ref{fig:Evo__FUV=1000_ALPHA=1e-03_lam=3}). 
}
\label{fig:diagnostics.w.r.t.psi_Rc=100_lam=3}
\end{figure*}

\begin{figure*}[h]
\centering
\includegraphics[width=1.025\textwidth]{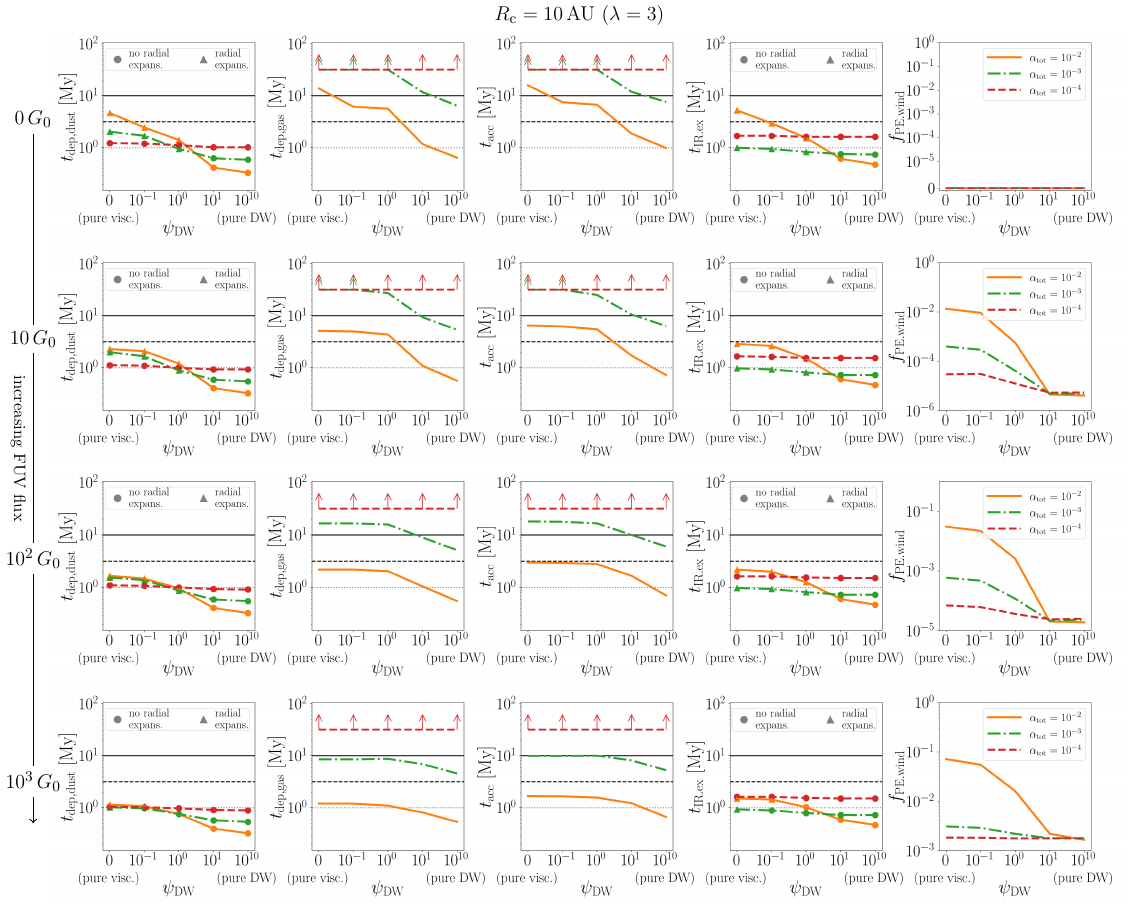}
\caption[]{Same as Fig.\ \ref{fig:diagnostics.w.r.t.psi_Rc=100_lam=3}, but for initially compact discs, $\Rc = 10\, \mathrm{AU}$.
Note again the somewhat inconsistent vertical axis range for the dust mass fraction removed by the external photoevaporative wind $f_\mathrm{PE.wind}$.}
\label{fig:diagnostics.w.r.t.psi_Rc=10_lam=3}
\end{figure*}

We use four diagnostics to extract disc lifetimes from our simulations.
We define $\tMg$ and $\tMd$ as, respectively, the times the gas mass falls below a given fraction $f$ of the initial gas mass (see purple vertical line of bottom row of Fig.\ \ref{fig:Evo__FUV=1000_ALPHA=1e-03_lam=3}), and the same for the (total) dust mass (see middle row of Fig.\ \ref{fig:Evo__FUV=1000_ALPHA=1e-03_lam=3}, red vertical lines). We take that fraction to be $f=1/1000$, as after this point the evolution happens fast enough that a change in $f$ does not lead to a significant change in  $\tMg$ or $\tMd$.
Then, $\tMacc$ is the time when the accretion rate onto the star falls below a given threshold, which we set to be $10^{-11}\,\mSun/\mathrm{yr}$.
Finally, $\tIRex$ is the time when the dust surface density of the \emph{small} grains at a reference radius of $\bar{R} = 1\,\mathrm{AU}$ is low enough that the disc becomes optically thin, so no IR excess would be detectable. We estimate the opacity of small grains to be $\kappa_0 \simeq 1000\,\mathrm{cm}^2/\mathrm{g}$ (e.g.\, \citealtads{1994A&A...291..943O}), and we thus check when $\Sigma_{\dust,0}(\bar{R})$ falls below $1/\kappa_0$.
We additionally track the fraction of dust mass removed by the photoevaporative wind $f_\mathrm{PE.wind}$ and the fraction of dust mass that is accreted onto the star $f_\mathrm{star}$.

The results for these diagnostics are shown in Figures \ref{fig:diagnostics.w.r.t.psi_Rc=100_lam=3} and \ref{fig:diagnostics.w.r.t.psi_Rc=10_lam=3} for various $\Rc$ and FUV fluxes, as a function of $\psiDW=\alphaDW/\alphaturb$ and $\alpha_\mathrm{tot}$ (see also Figs.\ \ref{fig:diagnostics_Rc=100_lam=3}, \ref{fig:diagnostics_Rc=30_lam=3}, and \ref{fig:diagnostics_Rc=10_lam=3}).
As we would expect, all timescales are shorter with higher external FUV fluxes, with a clear monotonic dependence (except for small $\Rc$ and high $\psiDW$, see below). The short lifetimes of discs under high FUV fluxes coupled with high accretion rates is a known problem (so-called `proplyd lifetime problem' due to the high mass-loss rates observed, e.g.\ \citealtads{1999AJ....118.2350H}, \citealtads{1999ApJ...515..669S}, \citealtads{2001ARA&A..39...99O}, and more recently e.g.\ \citealtads{2019MNRAS.490.5478W}).
This raises the question of whether the angular momentum transport mechanism within the disc plays a role, and naively one would expect that DW discs, since they don't spread out, would be less susceptible to the effects of external FUV irradiation and thus possibly live longer. 

In fact, we typically find that the \emph{opposite} is true. Because the driving factor to dust evolution is radial drift, the main difference does come from the ability of the disc to spread out due to diffusion. Like in Figure \ref{fig:Evo__FUV=1000_ALPHA=1e-03_lam=3}, we indicate those discs whose dust radii are able to initially expand with upward triangles in the first and fourth columns of Figures \ref{fig:diagnostics.w.r.t.psi_Rc=100_lam=3}, \ref{fig:diagnostics.w.r.t.psi_Rc=10_lam=3} while those that do not are labelled with circles. Strikingly, however, this implies that dust in viscous discs actually tends to live longer than in purely DW disc. Indeed, we find that, although viscosity can replenish the outer regions thus making VE discs more prone to losing dust via FUV winds (Fig.\ \ref{fig:Evo__FUV=1000_ALPHA=1e-03_lam=3}, middle row), this effect is completely overridden by the overall efficacy of dust radial drift driving accretion onto the central star (via the combined effect of the various processes mentioned at the beginning of Sect.\ \ref{sec:Results}): the dust in non radially-expanding discs has less of a radial distance to cover, and is thus removed more quickly. We therefore conclude that MHD winds do not protect the dust component \emph{of smooth discs} from being eroded by moderate-to-strong external FUV environments, and in fact the opposite is true. 

\emph{Results for $\tMd$.} Higher FUV fluxes also muddle the distinction between low and high $\alpha_\mathrm{tot}$, where a difference of this parameter of 2 orders of magnitude only leads to a difference in $\tMd$ of a factor of a few.
In light of this, initially larger discs, which are almost completely eroded from the outside in by stronger FUV fluxes, show a flat $\tMd$ dependence on $\psiDW$. Instead, initially very compact discs may still show signs of radial expansion even in very high FUV environments, at least for the highest $\alpha_\mathrm{tot}$. This means that some $\psiDW$ dependence in $\tMd$ is possible, but only for high $\alpha_\mathrm{tot}$, with $\tMd$ being a factor of a few \emph{lower} for DW discs than for VE discs. 
This stronger dependence of $\tMd$ on $\psiDW$ (as well as on $\alpha_\mathrm{tot}$) in initially smaller discs is expected as they will remain small if they are DW-driven, but not if they are viscously-driven; then, smooth dusty DW discs will tend to live shorter lives because they quickly accrete all the dust onto the star.
In higher FUV environments, $\alpha_\mathrm{tot}$ does not matter for $\psiDW\ll1$ (VE discs) while it does for high $\psiDW$ (DW discs). In this latter case (because, again, DW discs don't spread outwards), external photo-evaporation actually plays a minor role and the behaviour is similar to what we would obtain at $0\,\Gnaught$. We also remark that, when the external FUV flux is very low, the $\alpha_\mathrm{tot}$-dependence of $\tMd$ flips from correlated to anti-correlated for fixed $\psiDW$: $\tMd$ increases with $\alpha$ when $\psiDW=0$, but decreases with $\alpha$ when $\psiDW\gg1$.
That said, we should keep in mind that differences of only a factor of a few in disc lifetimes may be hard to discern observationally due to the difficulty of accurately determining YSO ages. This would be especially cumbersome for high FUV environments, as they tend to be farther away. 

\emph{Results for $\tMg$.} The gas depletion timescale follows a trend similar to $\tMd$. Discs typically tend to live longer for lower $\psiDW$, while this distinction is obfuscated by high FUV environments, especially for initially large discs. A noticeable difference with the dust is that the dependence on the total $\alpha_\mathrm{tot}$ remains instead very noticeable across all simulations. 
Moreover, except for the highest $\alpha_\mathrm{tot}$, $\tMg$ tends to be much larger than $\tMd$. This is again a consequence of dust radial drift.
We recall again that the evolution of gas radii has been already studied e.g.\ in \citeads{2024MNRAS.527.7588C}, which already pointed out how the evolution of radii can be indistinguishable in the DW and VE cases under high FUV irradiation (although they did not explicitly explore the dependence on $\Rc$).
$\tMacc$ shows clear correlation with $\tMg$, which is expected as the local mass flux is proportional to $\Sigma_\gas$, \eqref{eq:basicMHDAccrRate.parametrised.vDW}, with $\tMacc \gtrsim \tMg$. For purely viscously evolving discs, using $\tMacc \sim M_\disc/\dot{M}_\mathrm{acc}$, the need for high $\alphaturb$ to match the median gas age of $\simeq 3-5\,\mathrm{My}$ is in line with the classic results of \citeads{1998ApJ...495..385H,2000prpl.conf..377C}.

\emph{Results for $\tIRex$.} Concerning $\tIRex$, we see that trends with $\psiDW$ are similar to $\tMd$: larger $\psiDW$ lead to shorter $\tIRex$, with this difference being especially noticeable for those discs that are able to spread outward. Because $\tIRex$ tracks small dust, and most of the mass is in the large grains, we typically have that $\tIRex < \tMd$. 
We also note that the dependence on $\alpha_\mathrm{tot}$ is less straightforward for $\tIRex$ than in $\tMd$. While $\tMd$ shows a rather clear monotonicity with $\alpha_\mathrm{tot}$ (for a given fixed $\psiDW$), with higher $\alpha_\mathrm{tot}$ typically leading to a smaller $\tMd$ (except again for vanishing $\psiDW$ -- i.e.\ VE discs -- and small $\Rc$ as noted above, again due to disc spreading) this is not so for $\tIRex$, which is typically lowest for intermediate $\alpha_\mathrm{tot}$. 
The only exception to this is found for initially small discs and for $\psiDW\gtrsim 10$.

\emph{Results for $f_\mathrm{PE.wind}$.} Finally, looking at the dust mass fraction removed from the external photoevaporative wind, we find as expected (cfr.\ Fig.\ \ref{fig:f_wind.w.r.t.Rc} and subsect.\ \ref{subsec:DustEvo}) that $f_\mathrm{PE.wind}$ is higher for higher $\alphaturb$ but it is insensitive to $\alphaDW$ since wind-driven advection does not replenish the outer regions with dust material (as one can most easily see from the limiting cases $\psiDW\ll1$, i.e.\ VE discs, and $\psiDW\gg1$, DW discs). For low total $\alpha_\mathrm{tot}$, this dependence with $\psiDW$ is less severe, as even pure VE discs with low levels of turbulence have a hard time feeding dust into the outer regions.
Moreover, at fixed $\psiDW\lesssim10$, the increase of $f_\mathrm{PE.wind}$ with $\alphaturb$ is somewhat lessened for initially large discs once $\alphaturb\lesssim10^{-3}$, since the disc is already large enough and photoevaporation efficiencies will be rather similar. Indeed we also find that, as expected, higher FUV fluxes will lead to larger $f_\mathrm{PE.wind}$ (please note the different vertical scales in the last columns of Figs.\ \ref{fig:diagnostics.w.r.t.psi_Rc=100_lam=3} and \ref{fig:diagnostics.w.r.t.psi_Rc=10_lam=3} to accommodate for very different $f_\mathrm{PE.wind}$ ranges across all setups). For a fixed $\alpha_\mathrm{tot}$, $f_\mathrm{PE.wind}$ typically decreases with $\psiDW$, until it becomes almost insensitive to it once $\psiDW\gtrsim10$ (except in one case, namely $\Rc = 100\, \mathrm{AU}$, $\mathrm{FUV} = 10\,\Gnaught$ where this trend is reversed, albeit by an extremely small amount). We stress again that, in this work, $f_\mathrm{PE.wind}$ only tracks the dust that is removed by the external FUV wind, and we are not tracking dust that may be removed by MHD-disc winds (cfr.\ Appendix \ref{apx:DWDustEntr_RD}).

\section{Discussion}\label{sec:Discussion}
One of the main findings of our work is that, unlike one may naively expect, the angular momentum transport mechanism plays a rather limited role in modulating the lifetimes of dusty discs exposed to higher ($\gtrsim100\,\Gnaught$) FUV fluxes, and, if anything, DW discs tend in fact to live less than VE discs, both in the gas and in the dust. While the lifetime of the gaseous component still depends on the total $\alpha_\mathrm{tot} = \alphaturb + \alphaDW$, the dust lifetimes are instead less sensitive to it. In any case, as one would expect, lower $\alpha_\mathrm{tot}$ lead to a slower evolution overall, and thus somewhat longer disc lifetimes.
As we showed in the previous section, this is mostly due to radial drift of dust, which falls onto the star rather quickly. The role of radial drift was already pointed out by \citeads{2020MNRAS.492.1279S} in the purely viscous case, who showed that it mitigates the external photoevaporation's role in raising the dust to gas ratio in the disc.
Since higher $\psiDW$ values do not help, one may turn to $\alpha_\mathrm{tot}$. In principle, small $\alpha_\mathrm{tot}$ values do help in keeping dusty discs around for longer, at least for mild FUV fluxes, but as we noted above, the difference is rather small for more highly irradiated discs (FUV flux $\gtrsim 100\,\Gnaught$, bottom rows of Figs. \ref{fig:diagnostics.w.r.t.psi_Rc=100_lam=3}, \ref{fig:diagnostics.w.r.t.psi_Rc=10_lam=3}).
This points to the role of processes in the disc that can halt the inward flux of solids, most notable substructures (\citealtads{1972fpp..conf..211W}, \citealtads{2012A&A...538A.114P}). For purely viscous discs, dust traps can be able to extend the dust lifetime for FUV fluxes up to $\lesssim10^3\, \Gnaught$ as long as they are located inside the truncation radius \citepads{2024A&A...681A..84G}. Substructures are often observed in nearby discs in low-$\Gnaught$ environments (\citealtads{2015ApJ...808L...3A}, \citealtads{2016ApJ...820L..40A}, \citeyearads{2018ApJ...869L..41A}, \citealtads{2023ASPC..534..423B}, \citealtads{2025ApJ...989....9V}). Some discs experiencing moderate-to-high irradiation have also been observed. In $\sigma$ Orionis (FUV flux $\simeq 10^2 - 10^3\,\Gnaught$), \citeads{2024ApJ...976..132H} used the highest ALMA resolution possible and recovered substructures in 8 out of 8 discs in visibility modelling, with five of them have substructures unambiguously resolved in the image plane.
\citeads{2021ApJ...923..221O} observed 72 spatially resolved sources in the Orion Nebula, but with lower resolution, with some hints of the presence of substructures, although the bright large-scale emission makes it difficult to characterise disc morphology.
Our work suggests that, when demographic surveys are performed for discs in high-$\Gnaught$ environments, if we observe old discs with dust, they should also have substructures.

We stress again that this conclusion is largely independent of whether the discs' internal driving stresses are mainly viscous or MHD-driven. The question of telling the two mechanisms apart, as pointed out in the introduction, remains however a crucial one. In moderately-to-highly irradiated discs, it may unfortunately be challenging to directly disentangle hybrid discs from purely viscous discs. Indeed, as we showed in Section \ref{subsec:GasEvo}, hybrid and purely viscous discs display a rather similar correlations in the $\dot{M}_\mathrm{wind}$ v.s.\ $\dot{M}_\mathrm{acc}$ space, especially when thermally- and magnetically-driven winds are hard to distinguish themselves \citepads{2025arXiv251102811W}.

\subsection{Consequences for planet formation}
The removal of dust via external winds (large $f_\mathrm{PE.wind}$) implies that fewer grains are able to drift inward and replenish the inner regions. This would also have consequences for the chemical abundances (e.g., C/O, overall metallicity, etc.). Although highly irradiated regions are still poorly constrained by observations, recent JWST observations of the high-mass star-forming region NGC 6357 (\citealtads{2023ApJ...958L..30R}, \citeyearads{2025A&A...701A.139R}) seem to suggest that discs may be able to retain the ingredients necessary for rocky planet formation even in high FUV environments. \citeads{2024A&A...691A..32N} underlined the role of efficient radial drift, which may imply that external FUV fluxes even up to $10^4\,\Gnaught$ do not cause significant differences in chemical abundances of the inner regions compared to non-irradiated discs. However, they did not model dust entrainment in the external photoevaporative wind. More recently, \citeads{2025ApJ...991...94C} used thermochemical modelling (but not including drift) to conclude that inner disc chemistry may be affected, but the impact would be noticeable only for highly irradiated discs ($\gtrsim 10^6\,\Gnaught$). We will extend the current model to include chemical abundances in gas and dust phases in a future paper.

Substructures are important locations for planet formation: as solids accumulate here \citepads{2012A&A...538A.114P}, the dust-to-gas ratio locally increases, allowing growth, raising the efficiency of the streaming instability \citepads{2005ApJ...620..459Y} thus causing rapid planetesimal formation \citeads{2007Natur.448.1022J}. On the other hand, substructures are also believed to be a signpost for ongoing planet formation. It is still unclear if the substructures we see are indeed caused by planets as a general rule (e.g.\ \citealtads{2023A&A...677A..82T}) and whether such planets would represent the progenitors of the giant population we observe today (e.g.\ \citealtads{2019MNRAS.486..453L}, \citealtads{2022A&A...663A.163M}). Although our work suggests that substructures may play a relevant role in highly irradiated regions regardless of the nature of the driving internal stresses, we are unable to draw definite conclusion at this stage as we did not explicitly include them in our model.

For smooth discs that are mainly DW-driven, dust may survive somewhat longer if $\alpha_\mathrm{tot}$ is low (e.g.,\ in initially large discs, by a factor of up to $\sim 10$ for $10\,\Gnaught$, or a factor of $4-5$ for $100\,\Gnaught$; cfr.\ Fig.\ \ref{fig:diagnostics.w.r.t.psi_Rc=100_lam=3}). Such low $\alphaturb$ environments would have implications for solid growth, with smaller turbulent velocities leading to larger grains and faster radial drift, as well as stronger mid-plane concentrations \citepads{2024ARA&A..62..157B}.  A thinner solids layer leads to higher pebble accretion efficiency (\citealtads{2010MNRAS.404..475J}, \citealtads{2010A&A...520A..43O}, \citealtads{2014A&A...572A.107L}), although lower $\alpha$ also leads to a lower pebble isolation mass \citepads{2018A&A...612A..30B}, that is, smaller planetary cores. Planet-disc interactions are also affected: the ability to sustain a positive corotation torques to counterbalance the (typically negative) differential Lindblad torque is quenched at low diffusivities due to saturation (\citealtads{2011MNRAS.410..293P}, \citealtads{2014prpl.conf..667B}, \citealtads{2017MNRAS.471.4917J}, \citealtads{2019MNRAS.484..728M}); eccentricity- and inclination-damping rates are also affected in low-viscosity discs (\citealtads{2023A&A...670A.148P}, \citeyearads{2024ApJ...967..111P}). Orbital migration plays a crucial role in the assembly of planetary systems, as it often leads to assembly of resonant chains (e.g.\ \citealtads{2007ApJ...654.1110T}, \citealtads{2008A&A...482..677C}, \citealtads{2014A&A...569A..56C}). Low turbulence also affects in a measurable way the excitation of these resonant states (e.g.\ \citealtads{2021A&A...656A.115H}) and thus their stability properties \citepads{2018CeMDA.130...54P}.

\subsection{Limitations and future work}
Our model paints an incomplete picture of disc evolution by making a number of simplifying assumptions. We considered an axi-symmetric and vanishingly thin disc where quantities are a function of radius only. We followed \citeads{2022MNRAS.512.2290T} and adopted constant $\alphaturb$ and $\alphaDW$ in time and radius. A constant $\alphaturb$ with radius is likely unphysical, as local turbulence levels strongly depend on the disc's ability to develop the MRI, itself a function of the ionisation level in the disc (\citealtads{1996ApJ...457..355G}, \citealtads{2002ApJ...577..534S}, \citealtads{2002MNRAS.329...18F}).
A constant $\alphaDW$ may also not be realistic (e.g., \citealtads{2016A&A...596A..74S} showed it can depend on the local surface density and temperature). We will investigate non-constant $\alpha$ models in subsequent work.
We are also ignoring some internal processes that can affect disc structure, which challenge the smooth discs assumption. Substructures may be caused by MHD winds themselves (\citealtads{2017MNRAS.468.3850S}, \citeyearads{2019MNRAS.484..107S}, \citealtads{2019A&A...625A.108R}, \citealtads{2020A&A...639A..95R}, \citealtads{2022MNRAS.516.2006H}). Internal photoevaporation (e.g.\ \citealtads{2007MNRAS.375..500A}, \citealtads{2011MNRAS.412...13O}, \citealtads{2017RSOS....470114E}, \citealtads{2019MNRAS.487..691P}) also plays a role in the evolution, dispersal and delivery of solids to the inner region of the disc (\citealtads{2022MNRAS.514.2315C}, \citealtads{2023A&A...674A.165W}, \citeyearads{2025arXiv251101972W}, \citealtads{2024A&A...691A..72L}, \citeyearads{2025A&A...700A..67L}). 
However, external photoevaporation is expected to influence disc evolution throughout its whole life, whereas internal photoevaporation mainly drives final dispersal over a short final phase \citepads{2001MNRAS.328..485C}.
In any case, any process that would drive the formation of substructure is likely to have a measurable impact, since it would impede the efficient radial drift of dust.
We also ignore other external processes that may be important and affect disc structure, such as ongoing infall (\citealtads{2005ApJ...622L..61P}, \citealtads{2008AJ....135.2380T}, \citealtads{2010A&A...520A..17K}, \citealtads{2024A&A...691A.169W}).

\section{Conclusions}\label{sec:Conclusions}
In this paper, we presented for the first time the evolution of both the gaseous and solid components in smooth discs that are viscously-driven, MHD-wind driven, or both (hybrid discs), including the effect of external photoevaporation.
The disc evolution model followed that of \citeads{2022MNRAS.512.2290T}, with the addition of external photoevaporation (\citealtads{2020MNRAS.492.1279S}, \citealtads{2018MNRAS.481..452H}, \citeyearads{2023MNRAS.526.4315H}, \citealtads{2025ApJ...989....8A}), while the dust was modelled as a two population of small and large grains following \citealtads{2012A&A...539A.148B} (see also \citealtads{2017MNRAS.469.3994B}).
We evolve discs around a solar-mass star with different initial sizes, and with different strengths of viscous and MHD-wind stresses, parametrised by two \citeads{1973A&A....24..337S}-like parameters $\alphaturb$ and $\alphaDW$. We vary the external FUV flux from 0 to $1000\,\Gnaught$, and track various dust and gas lifetimes as well as the fraction of dust mass removed by the photoevaporative wind $f_\mathrm{PE.wind}$.

We find that, in more highly irradiated discs, the dust depletion time $t_\mathrm{dep,dust}$ is rather insensitive to the ratio $\alphaDW/\alphaturb$, with discs mostly driven by MHD winds having somewhat shorter lifetimes. This is because, while viscous discs spread out and are subjected to higher mass-loss rates from external photoevaporation for longer, inward dust drift onto the star results in faster dust mass depletion in MHD-wind discs, especially for initially small discs with high ``total'' $\alpha_\mathrm{tot} = \alphaDW + \alphaturb$. The dependence of $t_\mathrm{dep,dust}$ on the total $\alpha_\mathrm{tot}$ is itself weakened in the highest FUV environments, with $t_\mathrm{dep,dust}$ being only slightly longer for lower $\alpha_\mathrm{tot}$. This shows that, in highly irradiated discs, the total and relative strengths of viscous and magnetic stresses play a rather limited role in determining the lifetime of the dust species. While smooth discs with lower $\alpha_\mathrm{tot}$ may be able to retain dust for somewhat longer, other physical processes may be more relevant to shaping the disc lifetimes in high FUV environments, such as substructures, which can halt the inward drift of solids (\citealtads{1972fpp..conf..211W}, \citealtads{2012A&A...538A.114P}).

The fraction $f_\mathrm{PE.wind}$ of dust mass removed by the photoevaporative wind depends naturally on the FUV flux and the initial disc radius. It also shows a dependence on $\alphaDW/\alphaturb$ across all FUV fluxes considered here, at least for discs that are initially small and/or for higher total $\alpha_\mathrm{tot}$. As a general rule, $f_\mathrm{PE.wind}$ is higher for viscous discs (low $\alphaDW/\alphaturb$), as these spread outwards and can replenish the outer disc regions. However, this effect is somewhat less prominent for discs that are initially already large, as they don't need to rely on diffusion to bring material outwards, and with total $\alpha_\mathrm{tot}\sim 10^{-4}$. This will have consequences for planet formation as a higher $f_\mathrm{PE.wind}$ restricts the availability of solids and the chemical inventory in the inner disc.

Future work will address some limitations of this investigation, including additional effects which can give rise to substructures (e.g.\ internal photoevaporation or growing planetary cores). We will also add the tracing of chemical species to track the volatile delivery in the inner-most regions of discs and make predictions about important observables such as C/O ratios and metallicity.

\begin{acknowledgements}
GP and GR acknowledge support from the European Union (ERC Starting Grant DiscEvol, project number 101039651) and from Fondazione Cariplo, grant No. 2022-1217. RA acknowledges funding by the European Union through the E-BEANS ERC project (grant number 100117693). Neither the European Union nor the granting authority can be held responsible for them.
GL acknowledges support by PRIN-MUR 20228JPA3A and by the European Union Next Generation EU, CUP: G53D23000870006.
Views and opinions expressed are, however, those of the author(s) only and do not necessarily reflect those of the European Union or the European Research Council. 
\end{acknowledgements}

\bibliographystyle{aa}
\bibliography{references}

\begin{appendix}

\section{Comparison of photoevaporative models}\label{apx:FRIEDv1.vs.v2}
\begin{figure*}
\resizebox{\hsize}{!}
        {\includegraphics[height=0.4 \textwidth]{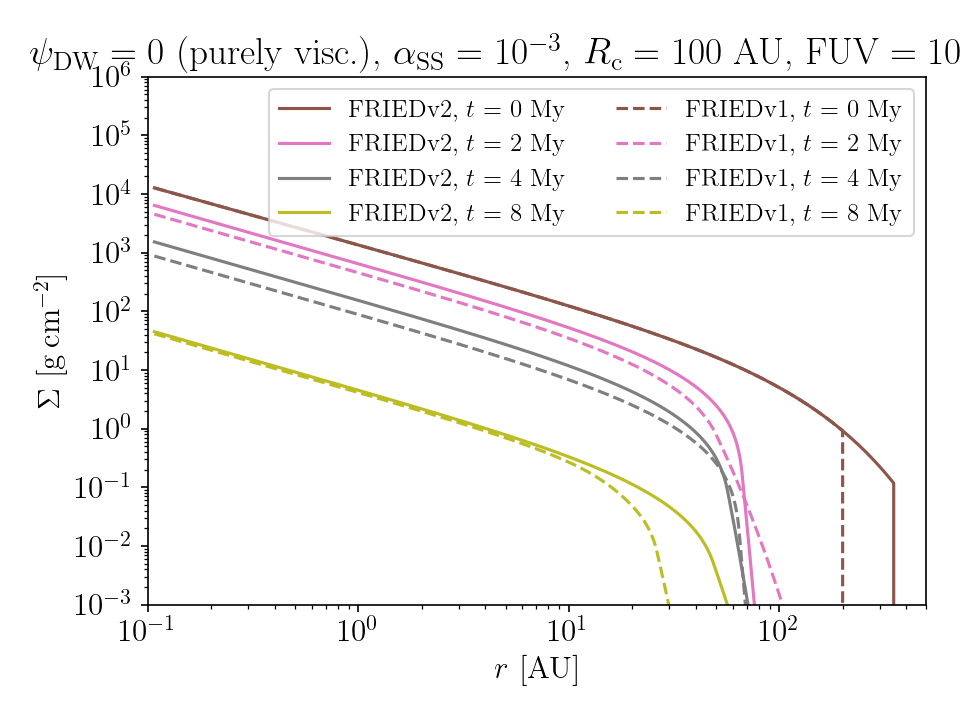}%
         \includegraphics[height=0.4125 \textwidth]{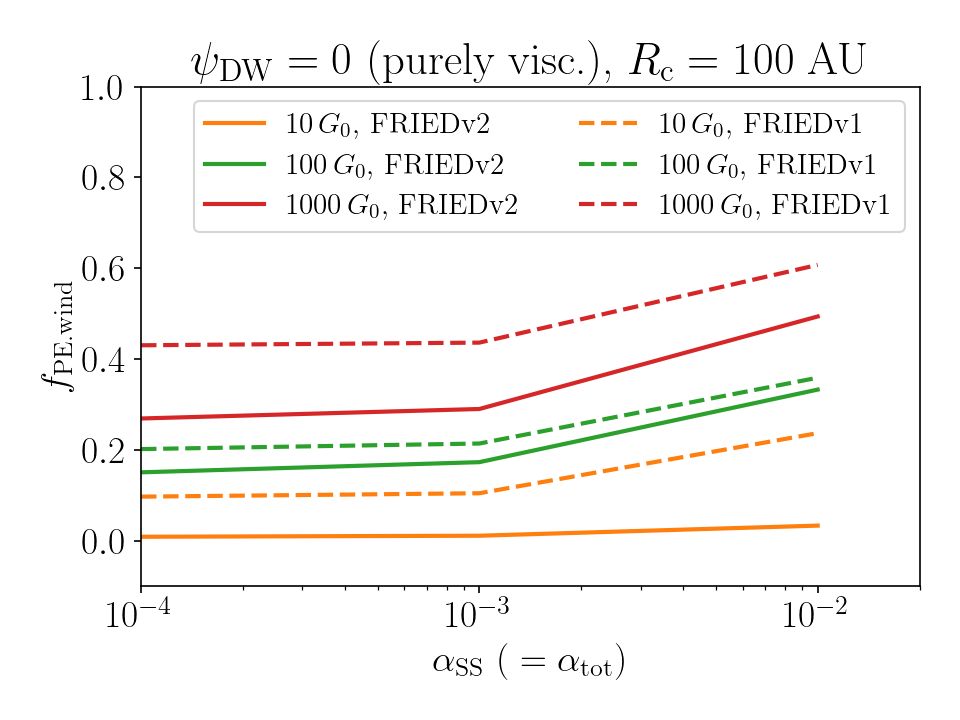}}
  \caption{Comparison of identical setups using the fiducial original FRIED grid (FRIEDv2) vs.\ the fiducial FRIEDv2 for the photoevaporative mass loss, in the case of purely viscous discs. \textit{Left panel}. Surface density evolution, for given $\alphaturb$ and FUV field. \textit{Right panel}. Fraction of dust mass loss to the wind for various $\alphaturb$ and FUV fields.
          }
     \label{fig:v1.vs.v2}
\end{figure*}

For honest comparison with previous works which used the original FRIED grid of \citeads{2018MNRAS.481..452H} (FRIEDv1), we show the outcome of a few setups with the original grid and with the updated version from \citeads{2023MNRAS.526.4315H} (FRIEDv2) we have used throughout the paper. The differences in the mass-loss rates arise from uncertainties in the wind microphysics, specifically the PAH-to-gas ratio $f_\mathrm{PAH}$ (more conservative in the original FRIED grid), the UV cross section of the wind (larger by about a factor 2 in the FRIEDv2 grid), and the dust-to-gas mass ratio in the wind (lower by a factor 3 in the updated grid), see Sect.\ 4 of \citeads{2023MNRAS.526.4315H}. 
The left panel of Figure \ref{fig:v1.vs.v2} shows the surface density evolution for purely viscous discs with $\alphaturb=10^{-3}$, $\Rc=100\,\mathrm{AU}$ and an FUV flux of $1000\,\Gnaught$, using the FRIEDv1 and FRIEDv2 grids. The right panel shows the dust mass fraction removed by the FUV-driven wind for different values of $\alphaturb$ and FUV fluxes (cfr.\ Figure 8 of \citealtads{2020MNRAS.492.1279S}) with the two different FRIED grids.
There are some minor differences in the surface density evolution, specifically in the disc truncation radii (cfr.\ Subsect.\ \ref{subsec:ExtPhotEvap} for the definition), but only by a factor of about a few. The main difference is in $f_\mathrm{PE.wind}$, which can vary by up to one order of magnitude for the lowest FUV fluxes, with the updated FRIED grid yielding (for the fiducial $f_\mathrm{PAH}=1$ value) consistently lower $f_\mathrm{PE.wind}$.
This shows that the outer disc microphysics may have some impact on the dust reservoirs in the inner discs (see also \citealtads{2025MNRAS.539.1190C}). While there is no FRIEDv2 analogue of the original FRIED grid tables, we investigate different $f_\mathrm{PAH}$ values in a separate paper (Pichierri et al.\ \emph{in prep}).

\section{Dust entrainment in MHD winds and role of radial drift}\label{apx:DWDustEntr_RD}
\begin{figure*}[h]
\centering
\begin{subfigure}{1.\textwidth}
    \centering
    \includegraphics[width=\textwidth]{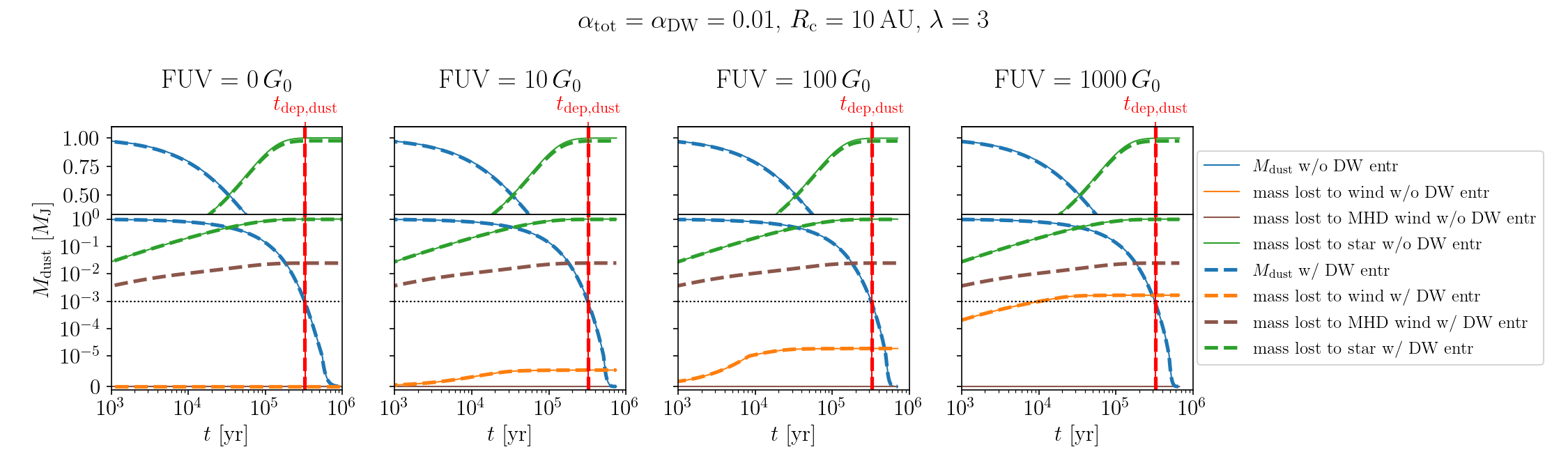}
\end{subfigure}
\begin{subfigure}{1.\textwidth}
    \centering
    \includegraphics[width=\textwidth]{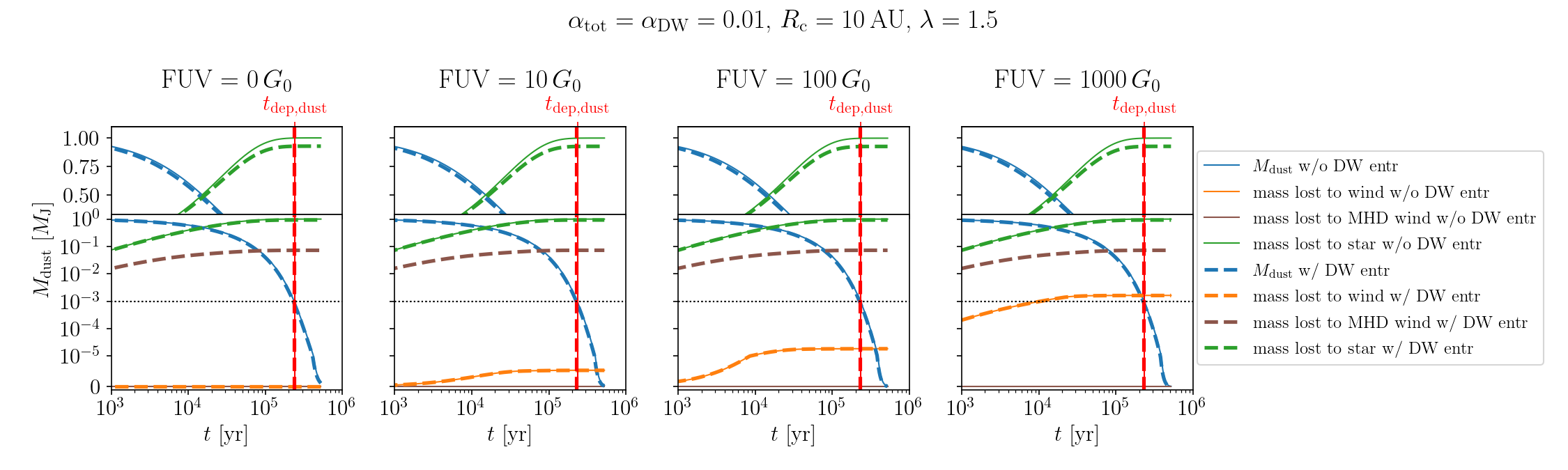}
\end{subfigure}
\caption[]{Evolution of the dust components in models including (thick dashed lines) or not including (thin continuous lines) dust entrainment and removal in MHD winds. 
The top panel represents the subset of the disc parameters used in this paper which would be most affected by MHD-wind dust entrainment, namely $\psiDW\gg1$ (MHD-wind discs), high $\alpha_\mathrm{tot} = \alphaDW = 0.01$, low $\Rc=10\,\mathrm{AU}$, our fiducial $\lambda=3$, and various FUV fluxes. We plot the evolution of the total dust mass, and the evolution of the dust mass lost to the photoevaporative wind, to the MHD wind and to the star (cfr.\ second row of Fig.\ \ref{fig:Evo__FUV=1000_ALPHA=1e-03_lam=3}). A top inset with a linear vertical axis scaling is added to better display the very small differences which would otherwise be undetectable.
The bottom panel shows the same disc setups but with $\lambda=1.5$, where MHD-wind-driven mass removal is most effective. Again, even in this extreme case, adding MHD-wind dust entrainment modifies only slightly the dust evolution and only affects the amount of dust accreted onto the star by a marginal amount (only visible in the top inset with linear vertical axis  scaling).}
\label{fig:DW_entr_PSI=1e10_ALPHA=0.01_RC=10}
\end{figure*}

We check that ignoring dust entrainment and removal through the MHD disc wind does not lead to significant differences to our results. The efficiency of this mechanism depends both on the ability of dust to be delivered to the wind and the ability of the wind to carry away dust of a given size. We follow \citeads{2021MNRAS.502.1569B}, which found that the former process is typically the limiting factor to dust removal, and that delivery of small grains to the wind base is induced by advection rather than turbulent diffusion. Using the results of \citeads{2019ApJ...882...33G} on the maximum size of dust than can be entrained by the MHD wind gives rather similar results within a factor of a few. Since the maximum grain size entrained scales with the local \emph{gas} mass-loss rate, from Eq.\ \eqref{eq:dotSigmagDW} the biggest difference in our case would occur for small discs with high $\psiDW$ and high $\alphaDW$; the mass-loss rate could in principle be higher than in our fiducial $\lambda=3$ if one assumes the extreme value $\lambda=1.5$, but the difference would only be of a factor of 4.
Figure \ref{fig:DW_entr_PSI=1e10_ALPHA=0.01_RC=10} shows a comparison in dust evolution for the representative limit cases of $\psiDW = \alphaDW = 0.01$, $\Rc=10\,\mathrm{AU}$ and $\lambda=3$ (top row), spanning different FUV fluxes. Only up to a few percent of the initial dust mass is removed by the MHD wind, which lowers the amount of dust accreted onto the star only slightly, but ultimately makes very little difference in the amount of mass removed by the external FUV wind, the evolution of total dust mass in the disc, and therefore the dust lifetime diagnostic $\tMd$.
Even considering the most extreme $\lambda=1.5$ (bottom row), the dust removed by the MHD wind consists at most of $\sim 10\%$ of the initial dust mass, which modifies the amount of dust that reaches the star by a small amount (see top plot which uses a linear vertical scale for clarity), but does not impact $f_\mathrm{PE.wind}$ or the total dust evolution significantly.

\begin{figure*}
    \includegraphics[width=1 \textwidth]{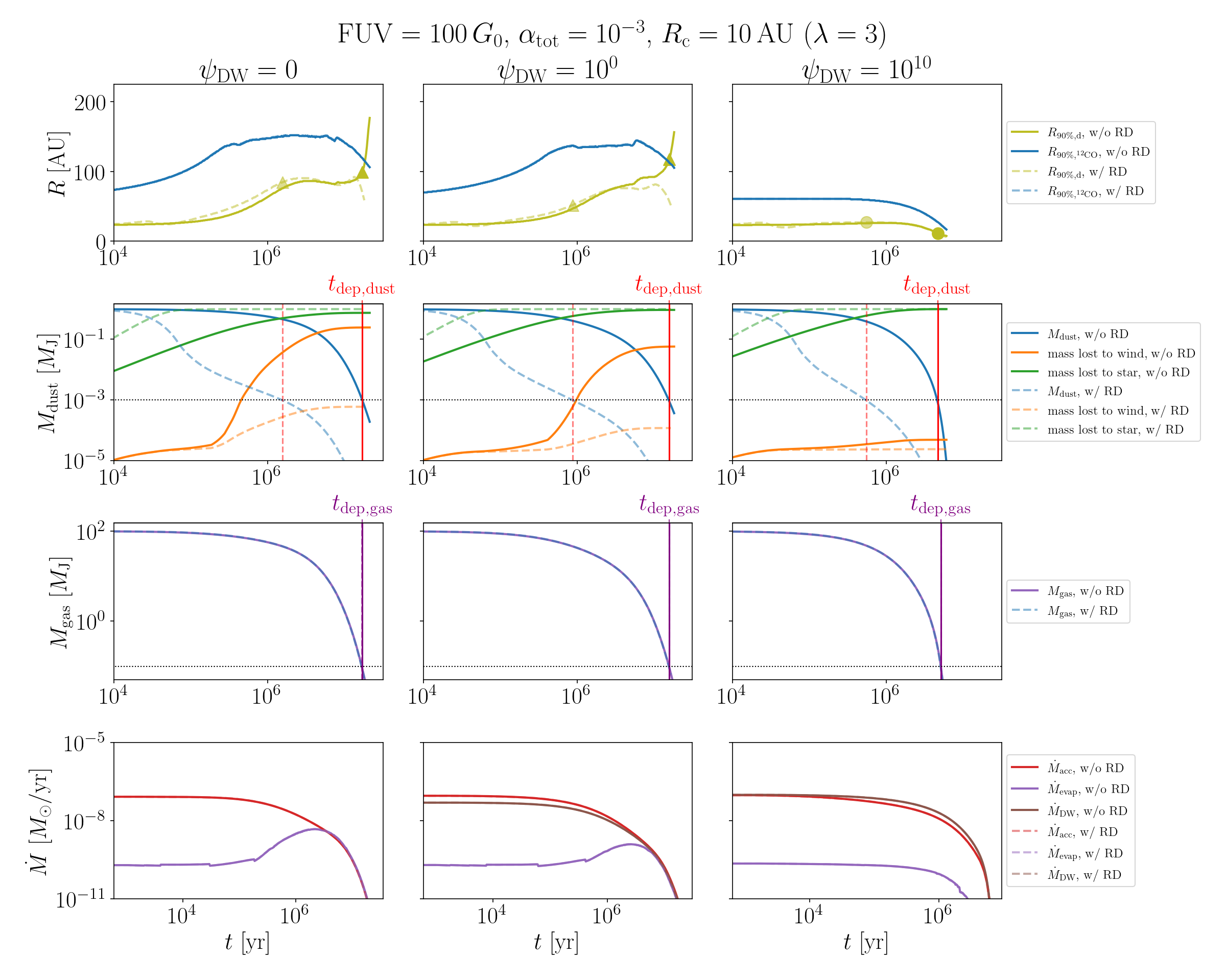}%
  \caption{Results for a setup including (dashed lines) and not including (continuous lines) dust radial drift. The top three rows plot the same quantities as in Fig.\ \ref{fig:Evo__FUV=1000_ALPHA=1e-03_lam=3}, with different columns corresponding to different $\psiDW = \alphaDW/\alphaturb$. We add the mass accretion rates as a bottom row (cfr.\ Fig.\ \ref{fig:dotMs_FUV=1000_lam=3}).}
     \label{fig:RD.vs.noRD}
\end{figure*}

We end this appendix by showing in Figure \ref{fig:RD.vs.noRD} how dust radial drift plays a decisive role in determining dust related diagnostics $\tMd$ and $f_\mathrm{PE.wind}$. While the evolution of the gas and the mass-loss rates are indistinguishable, the lack of radial drift severely impacts the dust lifetime and the amount of dust removed by the photoevaporative wind.  

\section{Single-fluid equations}\label{apx:CodeEqs}
We report for completeness the exact gas and dust evolution equations that are actually solved by the code (a version of the \lstinline{DiscEvolution} code described in \citealtads{2017MNRAS.469.3994B}). Numerically, we evolve the total surface density $\Sigma_\tot = \Sigma_\gas + \sum_k \Sigma_{\dust,k}$ solving an advection/diffusion equation with sink:
\begin{equation}
\begin{split}
    \frac{\partial \Sigma_\tot}{\partial t} & - \frac{3}{r}\frac{\partial }{\partial r}\left[\frac{1}{r\Omega_\Kepl} \frac{\partial}{\partial r}\left[\alphaturb c_\sound^2 r^2 \Sigma_\tot \right]
    \right] \\
    & - \frac{3}{2r}\frac{\partial }{\partial r}\left[\frac{1}{\Omega_\Kepl} \alpha_\mathrm{DW} c_\sound^2 \Sigma_\tot \right] \\
    & = s_\tot;\\
\end{split}
\end{equation}
the dust mass fractions $\epsilon_{\dust,k} = \Sigma_{\dust,k}/\Sigma_\tot$ evolves as
\begin{equation}
\begin{split}
    \frac{D\epsilon_{\dust,k}}{D t} &= -\frac{1}{\Sigma_\tot r}\frac{\partial}{\partial r}\left[r \Sigma_\tot \epsilon_{\dust,k}(\Delta v_{\dust,k} - \epsilon_{\dust} \Delta v_{\dust}) - r D_{\dust,k} \Sigma_\tot\frac{\partial \epsilon_{\dust,k}}{\partial r} \right] \\
    &~+ \epsilon_{\dust,k}\left(\frac{s_{\dust,k}}{\Sigma_{\dust,k}} - \frac{s_\tot}{\Sigma_\tot}\right).
\end{split}
\end{equation}
In the equations above, $s_\tot := s_\gas + \sum_k s_{\dust,k}$ represents the combined sink terms for gas and dust species, where in our case $s_\gas = -\left(\dot{\Sigma}_{\gas,\wind} + \dot{\Sigma}_{\gas,\extphot}\right)$, and $s_{\dust,k} = - \dot{\Sigma}_{\dust,k,\extphot}$ (no dust is removed by MHD winds in our nominal simulations presented in the main text for simplicity, although see Appendix \ref{apx:DWDustEntr_RD}, where a $-\dot{\Sigma}_{\dust,\wind}$ term is included). Moreover, $\frac{D}{D t} = \frac{\partial}{\partial t} + v \cdot \nabla$ is the co-moving derivative with the barycentric velocity $v$ of the different phases, $\Delta v_{\dust,k}$ is the relative velocity of the dust species $k$ to the gas (which takes into account the backreaction of dust onto the gas, \citealtads{2005ApJ...625..414T}, \citealtads{2010ApJ...722.1437B}), and $\epsilon_{\dust} \Delta v_{\dust} = \sum_k \epsilon_k \Delta{v}_{d,k}$. Following \citeads{2014MNRAS.444.1940L,2014MNRAS.440.2136L} (see also \citealtads{2017MNRAS.469.3994B}), this single-fluid approach is equivalent to a multi-fluid view of the problem.

\section{Explicit dependence on total and relative $\alpha$ parameters}
\begin{figure*}[h]
\centering
\includegraphics[width=1.025\textwidth]{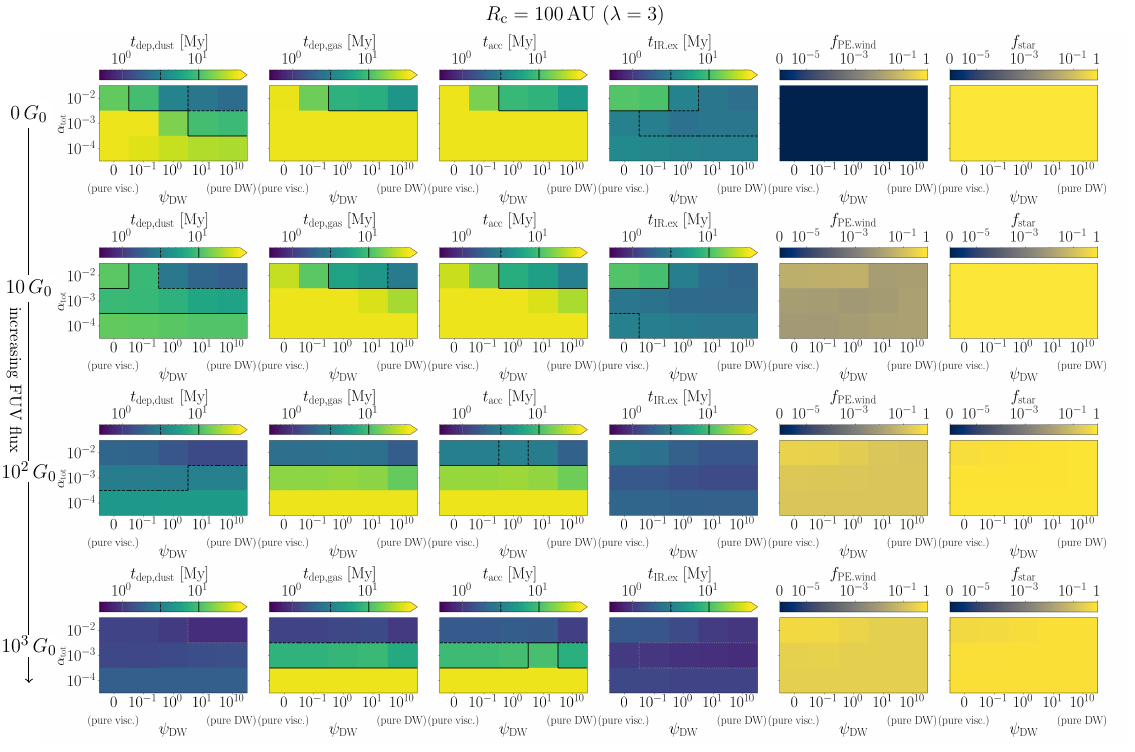}
\caption[]{
Similar to Fig.\ \ref{fig:diagnostics.w.r.t.psi_Rc=100_lam=3}, but displaying the dependence on $\alpha_\mathrm{tot}$ in the vertical axes, so each diagnostic is shown via a heat map. The fraction of dust mass accreted onto the star is also shown as an additional column on the right.
Regions of the $\psiDW$-vs-$\alpha_\mathrm{tot}$ parameter space where the timescales are above given thresholds ($1\,\mathrm{Myr}$, $3.16\,\mathrm{Myr}$ and $10\,\mathrm{Myr}$) are marked by dotted, dashed and continuous black lines (same scheme as Figs.\ \ref{fig:diagnostics.w.r.t.psi_Rc=100_lam=3}, \ref{fig:diagnostics.w.r.t.psi_Rc=10_lam=3})}
\label{fig:diagnostics_Rc=100_lam=3}
\end{figure*}

We show here an alternative visualisation of the data making Figures \ref{fig:diagnostics.w.r.t.psi_Rc=100_lam=3}, \ref{fig:diagnostics.w.r.t.psi_Rc=10_lam=3}, where the various diagnostics are plotted on the $\psiDW$-vs-$\alpha_\mathrm{tot}$ plane (Figs.\ \ref{fig:diagnostics_Rc=100_lam=3} and \ref{fig:diagnostics_Rc=10_lam=3}). This shows the regions of parameter space where the different disc lifetimes are above certain thresholds ($1\,\mathrm{Myr}$, $3.16\,\mathrm{Myr}$ and $10\,\mathrm{Myr}$). We also show both $f_\mathrm{PE.wind}$ and $f_\mathrm{star}$ on separate panels.

\begin{figure*}[h]
\centering
\includegraphics[width=1.025\textwidth]{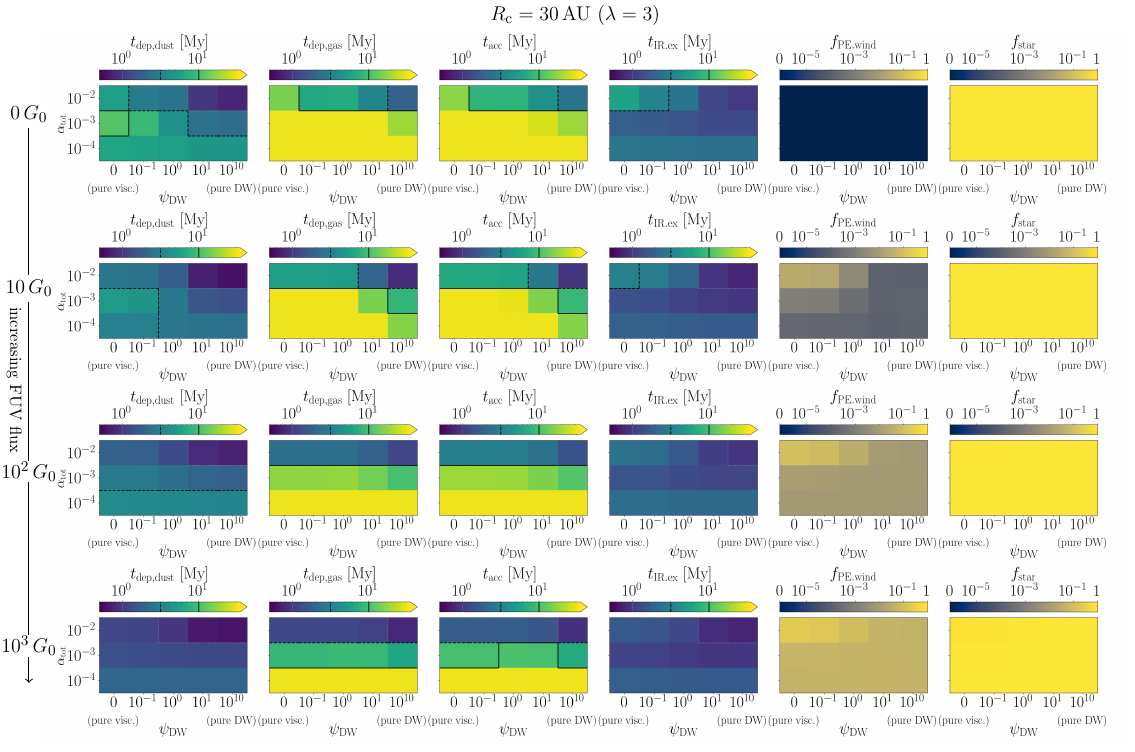}
\caption[]{Same as Fig.\ \ref{fig:diagnostics_Rc=100_lam=3}, but for initially less large discs, $\Rc = 30\, \mathrm{AU}$.}
\label{fig:diagnostics_Rc=30_lam=3}
\end{figure*}
\begin{figure*}[h]
\centering
\includegraphics[width=1.025\textwidth]{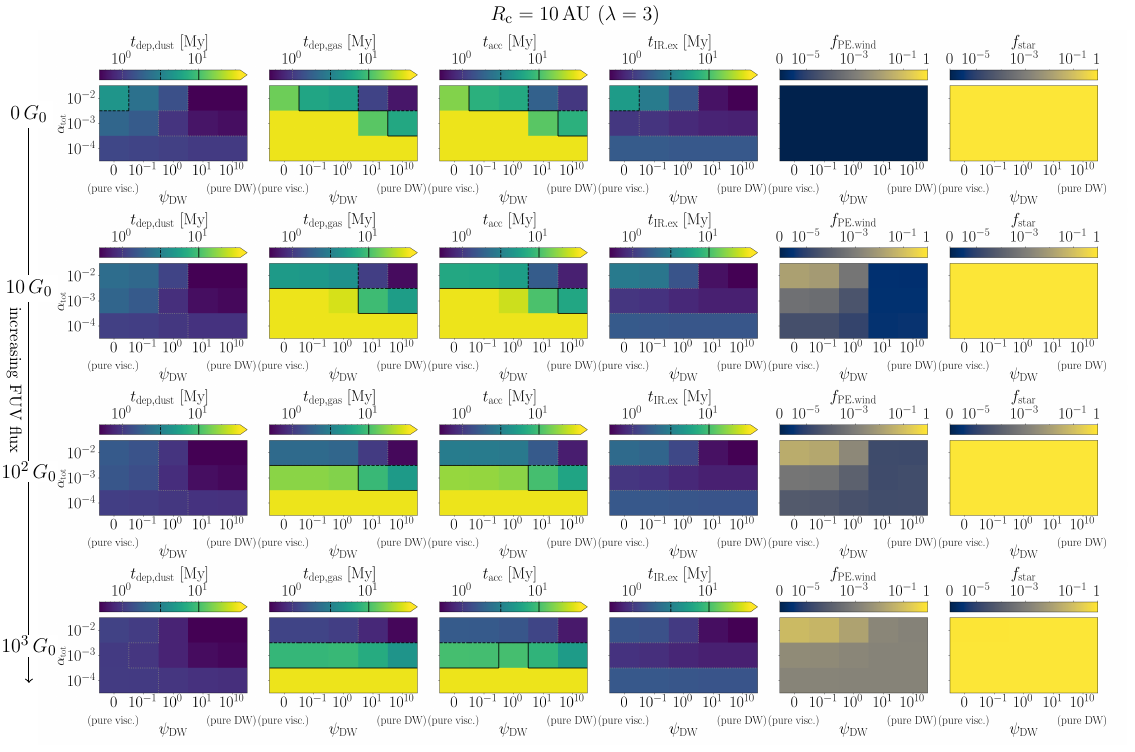}
\caption[]{Same as Fig.\ \ref{fig:diagnostics_Rc=100_lam=3}, but for initially compact discs, $\Rc = 10\, \mathrm{AU}$ (cfr.\ also Fig.\ \ref{fig:diagnostics.w.r.t.psi_Rc=10_lam=3}).}
\label{fig:diagnostics_Rc=10_lam=3}
\end{figure*}

\end{appendix}

\end{document}